\pdfoutput=1
\documentclass[preprintnumbers,amsmath,amssymb,prd,notitlepage,nofootinbib,twocolumn,superscriptaddress,table,xcdraw,floatfix]{revtex4-2}

\usepackage{graphicx}
\usepackage{dcolumn}
\usepackage{bm}
\usepackage{mathtools}
\usepackage{wasysym}
\usepackage{adjustbox}
\usepackage{ dsfont }
\usepackage{verbatim}
\usepackage[dvipsnames]{xcolor}
\usepackage{hyperref}
\usepackage{float}

\usepackage{aasmacros}
\usepackage{xspace}
\usepackage{tabularx}
\usepackage[font=footnotesize, caption=false]{subfig}
\renewcommand{\selectlanguage}[1]{}
\usepackage{blindtext}
\usepackage{verbatim}
\usepackage{mathrsfs}

\usepackage{tipa} 

\newcommand{\BibitemShut}[1]{} 

\usepackage{colortbl}
\mathchardef\mhyphen="2D

\makeatletter

    \def\CT@@do@color{%
      \global\let\CT@do@color\relax
            \@tempdima\wd\z@
            \advance\@tempdima\@tempdimb
            \advance\@tempdima\@tempdimc
    \advance\@tempdimb\tabcolsep
    \advance\@tempdimc\tabcolsep
    \advance\@tempdima2\tabcolsep
            \kern-\@tempdimb
            \leaders\vrule
                    \hskip\@tempdima\@plus  1fill
            \kern-\@tempdimc
            \hskip-\wd\z@ \@plus -1fill }
\makeatother

\renewcommand{\arraystretch}{1.2}

\date{\today}

\begin{document}

\title{Attention-based Neural Network Emulators for Multi-Probe Data Vectors Part III: 
Modeling The Next Generation Surveys}
\author{Yijie Zhu}
\affiliation{Department of Physics and Astronomy, Stony Brook University, Stony Brook, NY 11794, USA}

\author{Evan Saraivanov}
\affiliation{Department of Physics and Astronomy, Stony Brook University, Stony Brook, NY 11794, USA}
\affiliation{C. N. Yang Institute for Theoretical Physics, Stony Brook University, Stony Brook, NY 11794, USA}
\author{Joshua A. Kable}
\affiliation{C. N. Yang Institute for Theoretical Physics, Stony Brook University, Stony Brook, NY 11794, USA}
\author{Artemis Sofia Giannakopoulou}
\affiliation{Department of Physics and Astronomy, Stony Brook University, Stony Brook, NY 11794, USA}
\affiliation{C. N. Yang Institute for Theoretical Physics, Stony Brook University, Stony Brook, NY 11794, USA}
\author{Amritpal Nijjar}
\affiliation{Department of Physics and Astronomy, Stony Brook University, Stony Brook, NY 11794, USA}
\author{Vivian Miranda}
\affiliation{Department of Physics and Astronomy, Stony Brook University, Stony Brook, NY 11794, USA}
\affiliation{C. N. Yang Institute for Theoretical Physics, Stony Brook University, Stony Brook, NY 11794, USA}
\author{Marco Bonici}
\affiliation{Waterloo Centre for Astrophysics, ON, Canada N2L 3G1}
\author{Tim Eifler}
\affiliation{Department of Astronomy and Steward Observatory, University of Arizona, 933 N Cherry Ave, Tucson, AZ 85719, USA}
\affiliation{Department of Physics, University of Arizona,  1118 E Fourth Str, Tucson, AZ, 85721-0065, USA}
\author{Elisabeth Krause}
\affiliation{Department of Astronomy and Steward Observatory, University of Arizona, 933 N Cherry Ave, Tucson, AZ 85719, USA}
\affiliation{Department of Physics, University of Arizona,  1118 E Fourth Str, Tucson, AZ, 85721-0065, USA}

\begin{abstract}
Machine learning can accelerate cosmological inferences that involve many sequential evaluations of computationally expensive data vectors. Previous works in this series have examined how machine learning architectures impact emulator accuracy and training time for optical shear and galaxy clustering 2-point function. In this final manuscript, we explore neural network performance when emulating Cosmic Microwave Background temperature and polarization power spectra. We maximize the volume of applicability in the parameter space of our emulators within the standard $\Lambda$-cold-dark-matter model while ensuring that errors are below cosmic variance. Relative to standard multi-layer perceptron architectures, we find the dot-product-attention mechanism reduces the number of outliers among testing cosmologies, defined as the fraction of testing points with $\Delta \chi^2 > 0.2$ relative to \textsc{CAMB} outputs, for a wide range of training set sizes. Such precision enables attention-based emulators to be directly applied to real data without requiring any additional correction via importance sampling. Combined with pre-processing techniques and optimized activation and loss functions, attention-based models can meet the precision criteria set by current and future CMB and lensing experiments.  For each of Planck, Simons Observatory, CMB-S4, and CMB-HD, we find the fraction of outlier points to be less than $10\%$ with around $2\times10^5$ to $4\times10^5$ training data vectors. We further explore the applications of these methods to supernova distance, weak lensing, and galaxy clustering, as well as alternative architectures and pre-processing techniques. 
\end{abstract}
\maketitle

\section{Introduction}

Measurements of the Cosmic Microwave Background (CMB) from Simons Observatory \cite{SimonsObservatory:2018koc, SimonsObservatory:2025wwn}, CMB-S4 \cite{CMBS4:2021}, and CMB-HD \cite{Sehgal:2019ewc} are anticipated to provide unprecedented precision and resolution of both the temperature and polarization anisotropies. This leap in measurement capabilities will require corresponding improvements in the theoretical modeling. As such, the Boltzmann codes \textsc{CAMB} \cite{Lewis:1999bs} and \textsc{CLASS} \cite{CLASS/1:2011, CLASS-code} will have to compute the CMB TT, TE, and EE power spectra with higher accuracy settings. This, in turn, leads to a substantial increase in required computational resources and runtimes. 

This slowdown has direct implications for Markov Chain Monte Carlo (MCMC) runs when constraining cosmological parameters, with chains now needing days, sometimes weeks, to achieve appropriate convergence. Such high demand for computational resources will be further worsened in the exploration of models beyond $\Lambda$CDM  as they introduce additional computational complexity and increase the number of parameters \cite[see, e.g.][]{Reboucas:2024smm}.

Two main avenues exist to reduce the computational resources required to infer the probability distribution of cosmological parameters using Bayesian techniques, and they both can benefit from machine learning acceleration. First, there are innovative methods that sample the parameter space more efficiently, and they include Hamiltonian Monte Carlo (HMC)~\cite{Hajian:2006mt, Bonici:2023xjk, SPT-3G:2024atg,Piras:2023aub}, which is a more efficient method for sampling points during the MCMC process. In this case, deep learning can be used to approximate the probability distribution from a minimum set of MCMC accepted points \cite{Lange:2023ydq}. Another example is the nested sampling method~\cite{Feroz:2008xx}, which transforms the computation of the posterior to a one-dimensional integral over prior mass.

Alternatively, one can generate high-fidelity approximations for the outputs of Boltzmann codes, as shown in \citep{Kamionkowski/2021-no-boltzmann} where the authors propose to solve for the evolution of cosmological perturbations without the Boltzmann hierarchy. In this paper, we will emulate CAMB output for the TT, TE, and EE power spectra with machine-learning-based emulators \cite{cosmopower, Jense:2024llt}. Emulators offer several advantages over traditional MCMC, including the full parallelization of the computation of the training set.

Other emulators are capable of modeling the data vector space to parameter space, for instance, the neural density estimator explored in \cite{Wang:2023vej}. This method approximates the posterior by recurrently training emulators that go from data to parameters. In this manuscript, we do not consider this approach, as we would like to develop techniques that can be flexibly adapted to various analysis procedures more than MCMC, and produce products that are reusable in analyses involving varying scale cuts and so on.

The development of emulators for CMB power spectra is well established, and several works have trained machine learning models to emulate the CMB power spectra, including \textsc{CosmoPower}, which has been extensively validated on the multipole range and precision of both current and future experiments while covering moderate volumes in parameter space~\cite{cosmopower}. A recent work extended the multipole range in both standard and extended models for an expected allowed cosmological parameter volume from data from a Planck-like or more constraining CMB experiment~\cite{Bolliet:2023sst}. Furthermore, previous works have explored the impact of sophisticated pre-processing techniques~\cite{Bonici:2023xjk} and alternative machine learning architecture~\cite{Zhong:2024xuk, Saraivanov:2024soy}, on the performance of the emulator. 

Two aspects merit further analysis and motivate our work. The first pertains to the validation of machine learning emulators on cosmic-variance-limited experiments, with particular attention not only to the median likelihood error but also to the fraction of outliers with errors that exceed the acceptable threshold for a theoretical approximation to be applied without requiring additional refitting via importance sampling. The second aspect is related to the volume of parameter space that the emulator produces meaningful results in $\Lambda$CDM, as this can be a proxy for how the emulator scales with the number of parameters in extended models. The large volume of applicability also allows collaborations to apply blinding methods in their analysis by shifting the data vector by an arbitrary amount without risking generating nonsensical results. 

Motivated by these questions, we build a new class of attention-based CMB emulators, expanding two previous analyses that we hereafter refer to as part I \cite{Zhong:2024xuk} and part II \cite{Saraivanov:2024soy}. In Parts I and II, the authors applied so-called self-attention mechanisms, typically used in large language models, to emulate optical shear and clustering data vectors. They showed that these architectures reduce emulation error at a fixed training size. The attention mechanism was beneficial for mitigating outliers inconsistent with the maximum error criteria $\Delta \chi^2 = 0.2$ set by the Dark Energy Survey (DES) \cite{DES:2021rex}, which allowed for uniform emulator performance over a large parameter volume.

In Section~\ref{sec:ml}, we summarize the machine learning architectures, attention mechanisms, and activation functions that we explore in this work. In Section~\ref{sec:Training}, we outline our procedure for training emulators, and in Section~\ref{sec:results} we show results. In Section~\ref{sec:conclusion}, we provide discussions and conclusions. Finally, in the appendices, we present emulator performance for optical weak lensing, galaxy clustering, and uncalibrated supernova. 

\section{Machine Learning Models}\label{sec:ml}

\subsection{Architectures}\label{sec:arch}

Machine learning architectures define the data flow and the structural organization of the networks, which can significantly affect the emulator's performance given a set number of training points. The simplest ML architecture we consider in this work is the multi-layer perceptron (MLP), constructed by stacking blocks called dense layers formed by a linear transformation immediately followed by a nonlinear activation function. We assume that each linear transformation preserves the dimensionality of the data, simplifying the model by reducing the number of hyperparameters that can be tuned when selecting the final network architecture.

MLP designs can be prone to issues such as vanishing or exploding gradients \cite[see e.g.][]{Bengio/etal:1994, Razvan/etal:2012}. One effective strategy to alleviate this failure is the introduction of residual connections between groups of two or more dense layers \cite{2015arXiv151203385H}. A block formed by dense layers and a skip connection is called a ResMLP block, and a pure ResMLP architecture is composed solely of a series of ResMLP blocks. This notation closely follows the in-depth discussion presented in Part II \cite{Saraivanov:2024soy}.

The more advanced Transformer architecture differs from MLPs or ResMLPs by incorporating a self-attention layer that computes the similarity between different segments of the internal representation vector, potentially taking advantage of the correlated structure that exists in physically motivated data vectors. Inside the self-attention layer, the vector is subdivided into channels, and then an operation is performed to determine their similarity. Multiple valid options exist for such an operation, the original being a weighted-dot-product \cite{vaswani_attention_2017}. After the data vector passes through this attention layer, each channel is then further processed by residual dense blocks. The original implementation in Ref. \cite{vaswani_attention_2017} assumed identical ResMLP blocks on all channels, but in this work, they will be treated as independent. 

\begin{figure}[!ht]
    \centering
    \includegraphics[width=0.95\columnwidth]{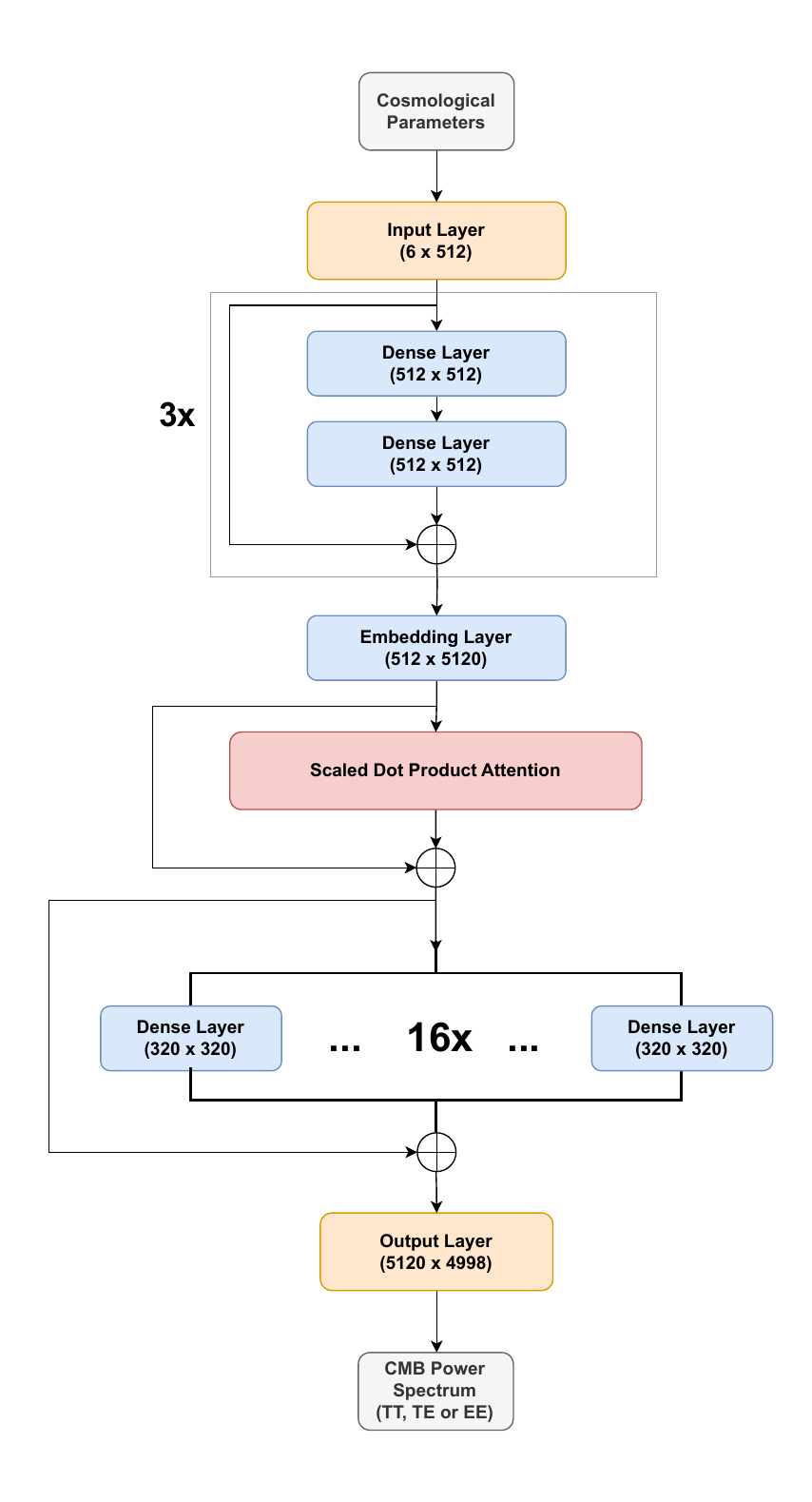}
    \caption{The baseline transformer architecture used in this work includes three ResBlocks and one 16-channel transformer block. The baseline transformer uses a dot product attention layer. We train an individual emulator for each of the TT, TE, and EE power spectra; however, we evaluate the accuracy of the emulators by concatenating all three power spectra together.}
    \label{fig:TRF}
\end{figure}

\subsection{Attention Mechanisms}\label{sec:att}
The original attention mechanism used in~\cite{vaswani_attention_2017,Zhong:2024xuk,Saraivanov:2024soy} is the dot-product attention. We apply this mechanism by first splitting the input data vector, $x$, into $N$ data vectors of length $d$, resulting in a $N \times d$ matrix, $X$. We transform each of these vectors using three different square weight matrices, $W_Q$, $W_K$, and $W_V$. This results in three separate $N \times d$ matrices, $Q=W_Q X$, $K= W_K X$, and $V=W_V X$. We then perform the following operation:
\begin{equation} \label{attention_eq}
    Y(Q,K,V) =\textsc{Softmax}\left(\frac{QK^T}{\sqrt{d}}\right)V\,,
\end{equation}
Where $QK^T$ is the $N\times N$ matrix of dot products between each vector in $Q$ and each vector in $K$. Each of these products acts as a weight to the vectors in $V$. The factor $\sqrt{d}$ is inserted to prevent the gradient from vanishing when the length of the data vector is too large~\cite{vaswani_attention_2017}. Furthermore, the \textsc{Softmax} function is defined as
\begin{equation} \label{softmax}
    \textsc{Softmax}(y_j)=\frac{e^{y_j}}{\sum_i e^{y_i}},
\end{equation}
where $i,j$ are subscripts of elements in the data vector. We use this dot product attention mechanism in our baseline emulator and show schematically the architecture in Figure~\ref{fig:TRF}.

\begin{table}
    \centering
    \renewcommand{\arraystretch}{1.3}
    \begin{tabular}{lcccc}\hline
        & Uniform & Uniform & Gaussian & Gaussian \\
        Parameter & Train & Test & Train & Test\\\hline
        $100\Omega_b h^2$  & $[0.6,3.8]$ & $[0.8,3.5]$ & $[0,4]$      & $[0,4]$ \\
        $10\Omega_c h^2$   & $[0.30,2.35]$  & $[0.40,2.2]$   & $[0,3]$       &  $[0,3]$ \\
        $H_0$              & $[25,114]$      & $[30,110]$      & $[25,114]$      & $[30,110]$\\
        $10\tau$           & $[0.07,1.5]$  & $[0.1,1.4]$   & $[0.07,1.5]$  & $[0.1,1.4]$ \\
        $\log(10^{10}A_s)$ & $[1.61,3.6]$    & $[1.7,3.5]$     & $[1.61,4.5]$    & $[1.7,4.0]$ \\
        $n_s$              & $[0.7,1.3]$     & $[0.8,1.2]$     & $[0.7,1.3]$     & $[0.8,1.2]$ \\\hline
    \end{tabular}%
    \caption{Parameter priors for the training and testing parameter sets for the uniform and Gaussian sampling methods.  }
    \label{tab:parameter_ranges}
\end{table}

In addition to the dot product attention mechanism, we test three alternative attention mechanisms that could allow more efficient training or better accuracy with the Fast-Transformer modules\cite{katharopoulos_et_al_2020,vyas_et_al_2020}. The first alternate mechanism we test is called the linear attention model \cite{linatt}, which is motivated by the first-order Taylor approximation of the exponential that appears in the \textsc{Softmax} function. This approximation assumes that the dot product $QK^{T}$ is small, which would occur if there are few common features in the data vector. In this case, plugging Equation~\ref{attention_eq} into Equation~\ref{softmax} results in exponential terms that to linear order are approximately
\begin{equation}
    e^{Q_iK_j^T/\sqrt{d}} \approx 1+\frac{Q_iK_j^T}{\sqrt{d}}\,,
\end{equation}
which reduces the computational resources for evaluating the attention matrix, $Y$. For the linear attention model, we replace all the exponential functions in Equation~\ref{softmax} with this approximation~\cite{linatt}.

\begin{table}[t]
\centering
\renewcommand{\arraystretch}{1.3}
\begin{tabular}{ccc}
    \hline
    Parameter            & {}\quad{} & Value \\\hline
    $\Omega_bh^2$        & & $0.02239$ \\
    $\Omega_ch^2$        & & $0.1178$ \\
    $H_0({\rm km/s/Mpc})$                & & $67.5$ \\
    $\tau$               & & $0.06$ \\
    $\log{(10^{10}A_s)}$ & & $3.064$ \\
    $n_s$                & & $0.965$ \\
    $\Sigma m_{\nu}({\rm eV})$     & & $0.06$ \\
    \hline
\end{tabular}
\caption{The fiducial cosmological parameters in flat $\Lambda$CDM we adopt throughout the manuscript. These values are chosen to be close to the Planck best-fit cosmology~\cite{Planck:2018vyg}. }
\label{tab:fiducial}
\end{table}

\begin{figure*}[t]
    \centering
    \includegraphics[width=0.94\textwidth]{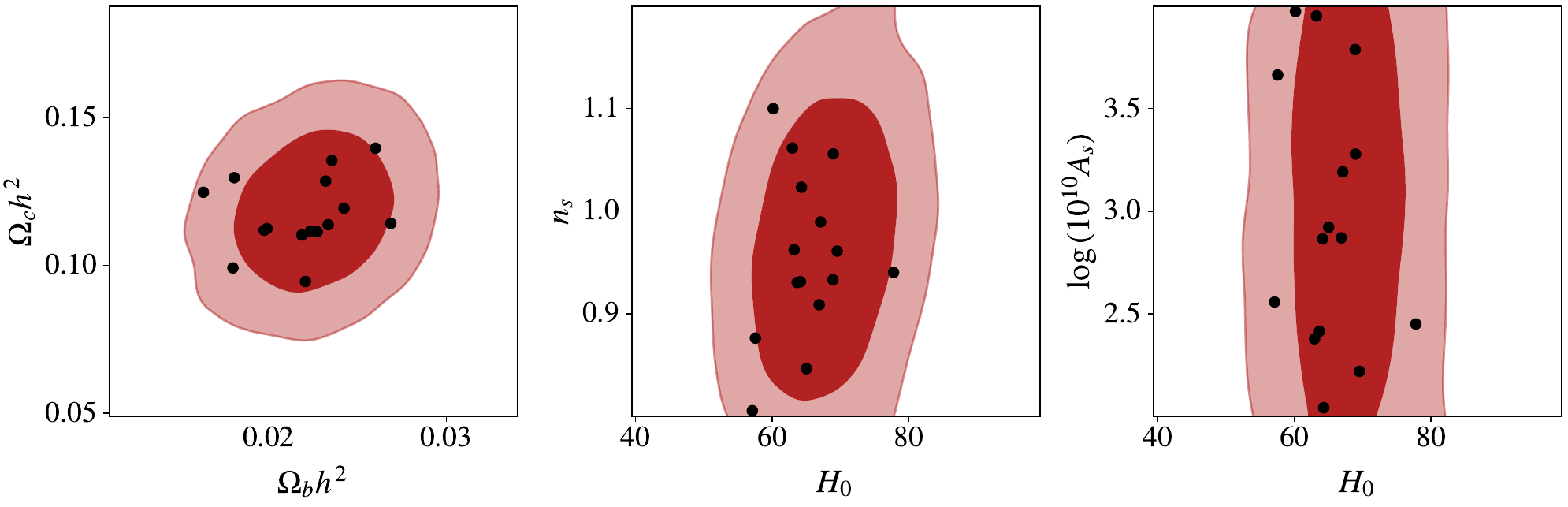}
    \includegraphics[width=0.3\textwidth]{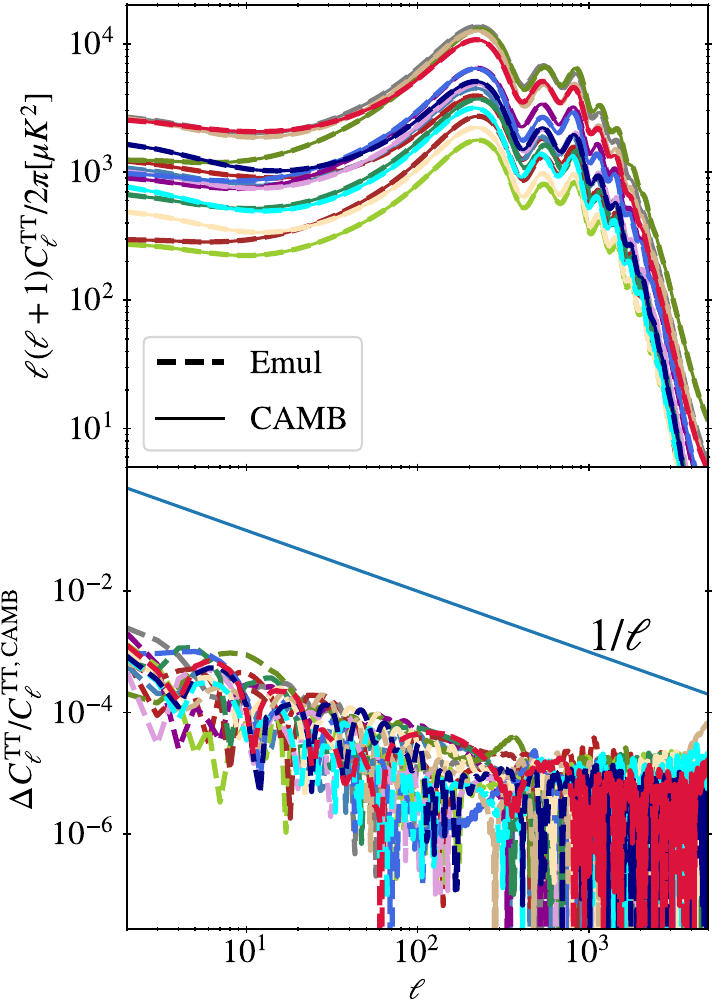}
    \includegraphics[width=0.3\textwidth]{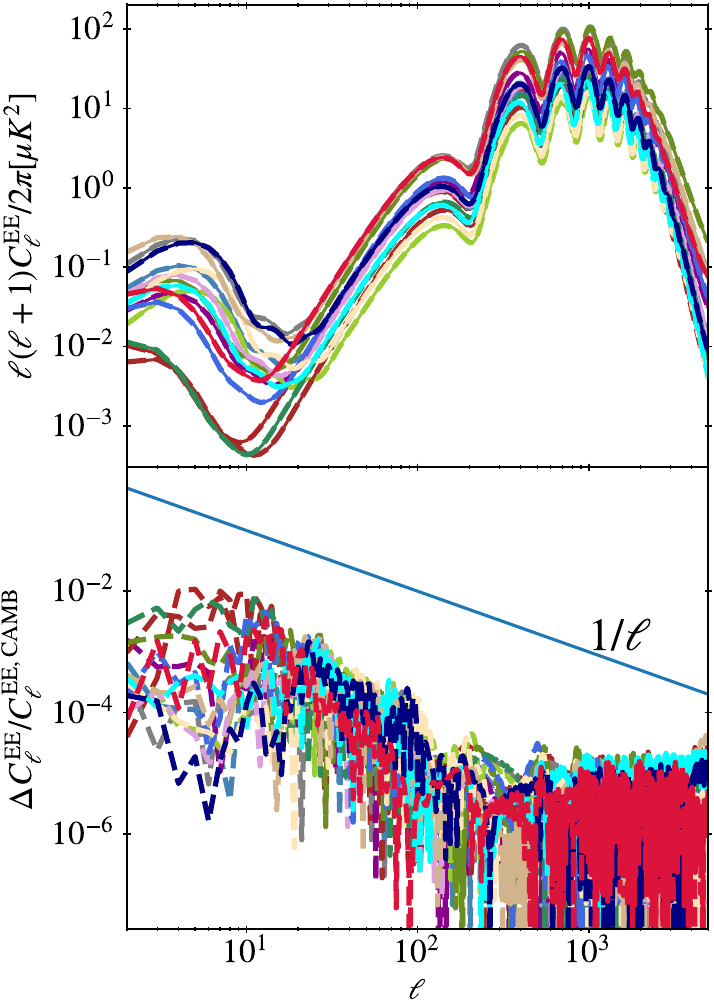}
    \includegraphics[width=0.3\textwidth]{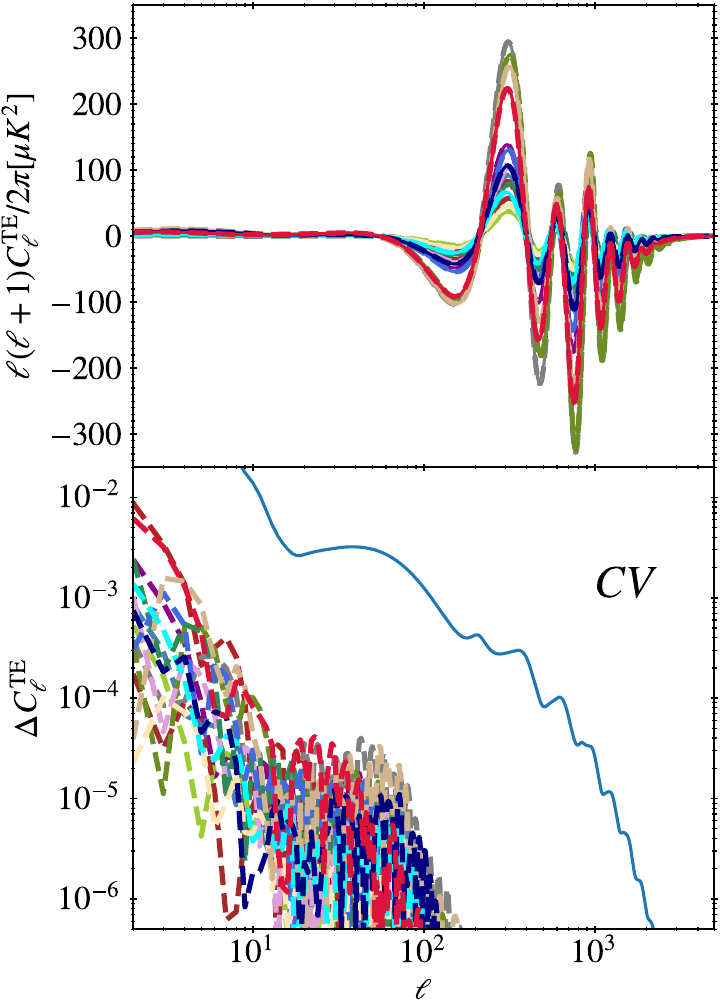}
    \caption{ Comparison between emulator outputs and \textbf{CAMB} outputs for $15$ different cosmologies. On the top panel, we show the positions of the testing points from the $T=128$ Gaussian sampling set that are used to calculate the TT, TE, and EE power spectra. For the emulation, we use our baseline transformer model trained on $530$ thousand training points, which is described in detail at the start of Section~\ref{sec:results}. From left to right, we show results for each of the TT, EE and TE spectra. The middle panels show direct comparison between emulator outputs and the true outputs. Lower panels show a comparison between $1/\ell$ scaling and the fractional difference between the emulator and \textbf{CAMB} outputs. We can see that, in general, the emulator errors are below the cosmic variance limits. }
\label{fig:range}
\end{figure*}

Similar to the idea of linear attention, latent attention also aims at reducing the computational cost of attention matrices by breaking the large attention matrices into decompositions of smaller ones \cite{latent}. The general idea is to write, for some large matrix $A$,
\begin{equation}
    A^{(n\times n)}\approx A^{'(n\times n)}=A_1^{(n\times k)}A_2^{(k\times n)}.
\end{equation}
The number of trainable parameters in $A^{(n\times n)}$ is $n^2$, while it is $2nk$ in the combination $A_1^{(n\times k)}$ and $A_2^{(k\times n)}$. As long as $k\leq n/2$, this approximation reduces the amount of trainable parameters needed. To test this, we replace all $K, Q, V$ matrices with this decomposition, and use $k=n/4$ for a factor of $2$ reduction of parameter number in the attention layers.

Alternatively, the data vector may have features that are important for the emulator to learn. The Locality Sensitive Hashing (LSH) attention model \cite{lsh} is designed to find these most important features while ignoring smaller common features in the data. The LSH model is designed to reduce the computational cost associated with the calculation of the $(QK^T)$ matrix, an intermediary, yet computationally intense, step in computing $\textsc{Softmax}(QK^T)$ in Equation~\ref{attention_eq}. Qualitatively, the LSH scheme only evaluates the larger elements of $QK^T$, reducing the number of rounds of evaluations for this $\mathcal{O}(d^2)$ computation. The $QK^T$ matrix is recast as a block diagonal matrix, where each block contains the dot product on a subset of the $Q$ and $K$ matrices that are most similar to one another. This similarity is determined by a random projection LSH algorithm that randomly and systematically partitions the space of the $K$ matrix into a number of hashing buckets, then defines parts of the $K$ and $Q$ matrices that fall into the same bucket as similar. The dot product is calculated only for the parts of the $K$ and $Q$ matrices that fall within the same hashing buckets, leading to a block diagonal matrix, so the computation uses fewer resources.
Finally, we explore the attention-free transformer (AFT) model \cite{aft}, which avoids the matrix multiplication in Equation~\ref{attention_eq} and instead uses an element-wise product. This reduces the number of computations that need to be performed as information passes through this layer, which could allow it to scale to larger data vectors more efficiently than the standard dot product mechanism. For the offsets not including the full matrix multiplication, the AFT adds an additional position bias matrix, $w$, of new trainable parameters that allows the model to learn long-range correlations in the data. In particular, the formula for calculating the attention-weighted output from the AFT model is
\begin{equation}
    Y_t=\sigma_{\rm{q}}(Q_t)\odot\frac{\sum_{t'=1}^N\exp{(K_{t'}+w_{t,t'})}\odot V_{t'}}{\sum_{t'=1}^N\exp{(K_{t'}+w_{t,t'})}},
\end{equation}
where $Q,K,V$ are defined the same way as in the dot-product attention, $\odot$ is element-wise multiplication, and $\sigma_{\rm{q}}$ is set to be a \textsc{Sigmoid} function.

\begin{figure*}[t]
    \centering
    \includegraphics[width=0.94\textwidth]{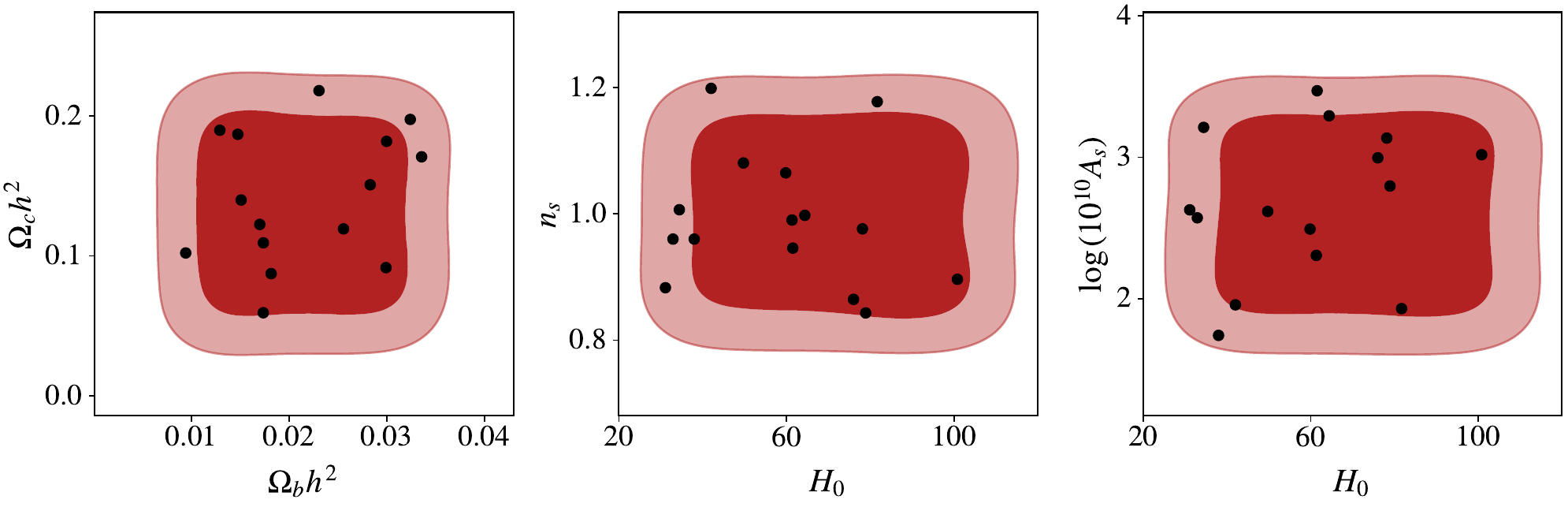}
    \includegraphics[width=0.3\textwidth]{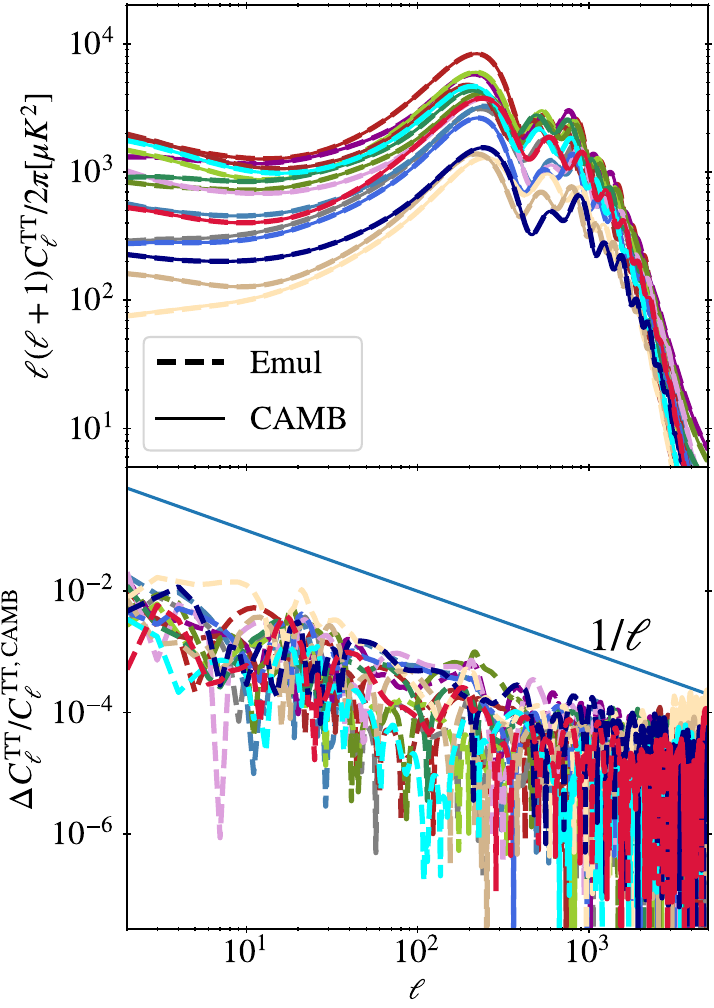}
    \includegraphics[width=0.3\textwidth]{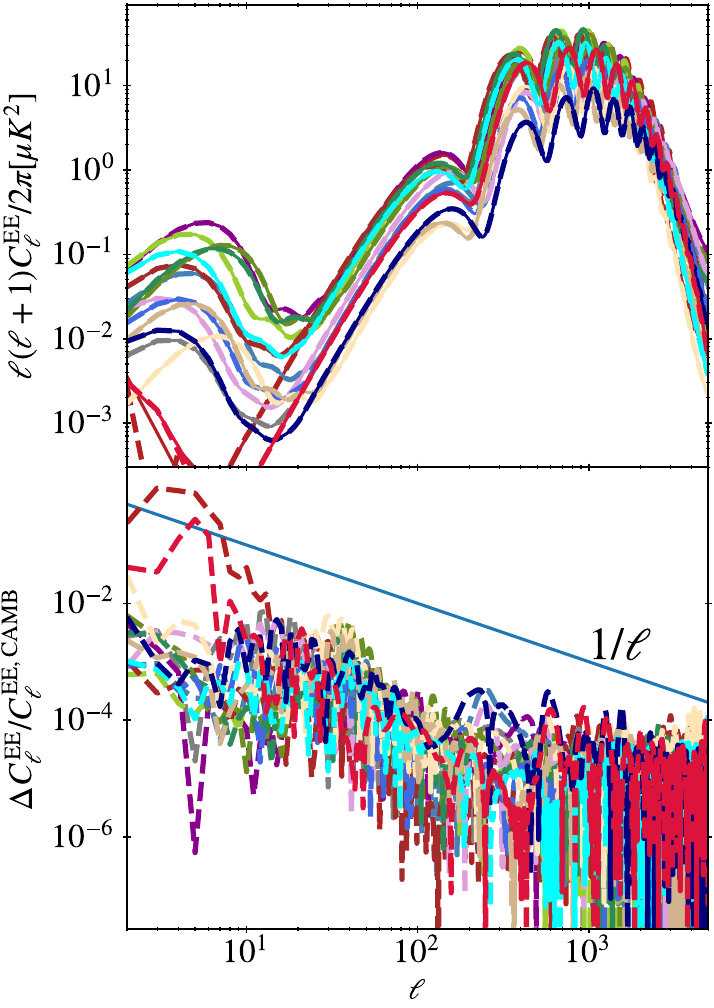}
    \includegraphics[width=0.3\textwidth]{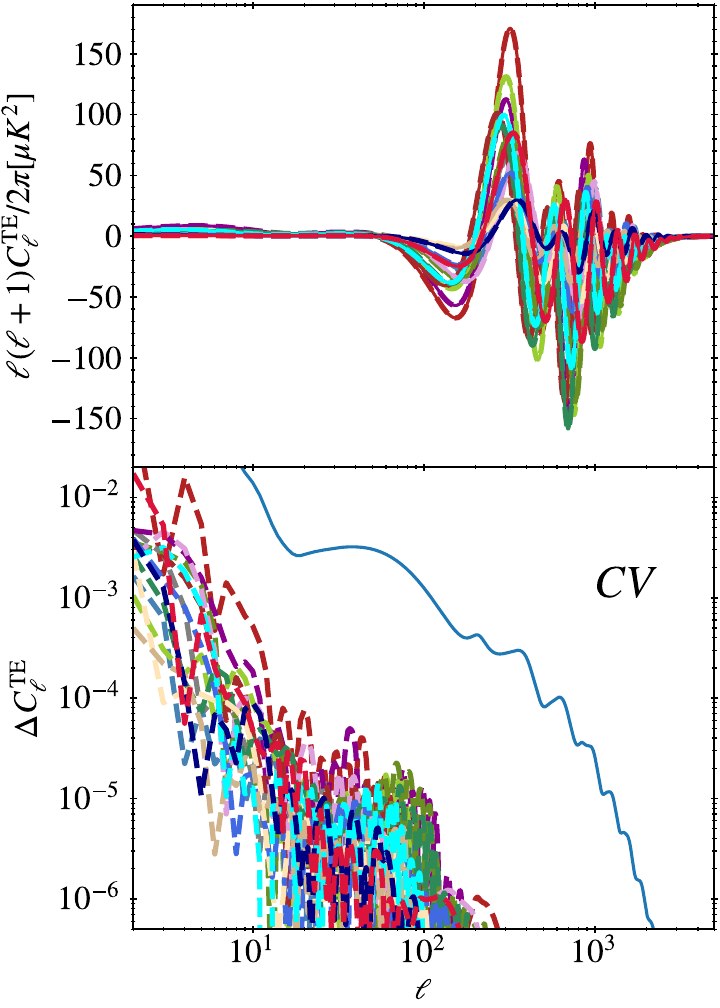}
    \caption{ Comparison between emulator outputs and \textbf{CAMB} outputs for $15$ different cosmologies. On the top panel, we show the positions of the testing points from the uniform sampling set that are used to calculate the TT, TE, and EE power spectra. The emulator model here is the baseline transformer model, which is described in detail at the start of Section~\ref{sec:results}, trained with  $530$ thousand training points. From left to right, we show results for each of the TT, EE and TE spectra. Upper panels show direct comparison between emulator outputs and the true outputs. Lower panels show the comparison between $1/\ell$ scaling and the fractional difference between the emulator and \textbf{CAMB} outputs. When trained and tested on the uniform sets, the emulators are able to produce approximated data vectors to high precision under cosmic variance limit.}
\label{fig:unirange}
\end{figure*}

\subsection{Activation Function}\label{sec:af}

Activation functions enable neural networks to learn nonlinear behavior. Among the most common choices is the Rectified Linear Unit (\textsc{ReLU}), which behaves linearly for positive inputs while returning zero for negative inputs. This behavior may not be optimal for normalized inputs, as we normalize both the input and output of the emulator by taking $\tilde{y}=(y-\langle y\rangle)/\sigma_y$ (i.e, both input and output of the model will be centered around $0$ with variance equal to $1$). This motivates using the alternative \textsc{Tanh} function as was explored in Part II \cite{Saraivanov:2024soy}, where the \textsc{Tanh} activation function was shown to perform better for weak lensing emulators. 

A third activation function, denoted  \textsc{h}(x) and previously used in~\cite{Alsing/etal:2020, cosmopower, CONNECT} is defined as 
\begin{multline}\label{eq:activation_fcn}
    \textsc{h}(x)=\Big(\gamma+(1+e^{-\beta\odot x})^{-1}\odot(1-\gamma)\Big)\odot x \, ,
\end{multline}
where $\gamma$ and $\beta$ are trainable vectors with the same shape as the data vector, $x$. The vector $\gamma$ determines the mixing between the linear and nonlinear components, while $\beta$ controls the sharpness of the function near $x=0$. According to~\cite{Alsing/etal:2020}, the flexibility of this activation function allows it to capture both sharp and smooth features that may be present in the data. 

\section{Training Process}\label{sec:Training}

This section outlines the adopted training strategies, sampling, and pre-processing methods used in this work. Their performance will be evaluated using two primary metrics:  by the predicted median, $\langle \Delta \chi^2 \rangle_{\rm med}$, and the fraction of outliers, \texthtbardotlessj$(\Delta\chi^2>0.2)$, in the testing sample, with $\Delta \chi^2$ defined as
\begin{equation} \label{eq:delta_chi2}
    \Delta \chi^2 \equiv \big(\boldsymbol{y}_{\textrm{CAMB}} - \boldsymbol{y}_{\textrm{emu}}\big)^T \ \Sigma^{-1} \ \big(\boldsymbol{y}_{\textrm{CAMB}} - \boldsymbol{y}_{\textrm{emu}}\big) \, .
\end{equation}
Here, $\boldsymbol{y}_{\textrm{Z}}$ is the data vector formed by the concatenation $\big(C_{\ell}^{\rm TT}, C_{\ell}^{\rm TE}, C_{\ell}^{\rm EE}\big)$, with $C_{\ell}^{\rm XY}$ computed either by \textrm{CAMB} or the machine-learning emulator. This data vector is the concatenation of the CMB temperature (TT) and polarization (EE) auto-spectra, as well as their (TE) cross-spectra. 
In addition, $\Sigma$ denotes the data covariance for a cosmic-variance-limited experiment
\begin{equation}\label{eq:cosmic_variance_covmat1}
\begin{split}
    \Sigma&=
      \begin{pmatrix}
        \Sigma^{\rm TTTT}_{\ell \ell'} & \Sigma^{\rm TTTE}_{\ell \ell'}  & \Sigma^{\rm TTEE}_{\ell \ell'}\\
        \Sigma^{\rm TTTE}_{\ell \ell'}  & \Sigma^{\rm TETE}_{\ell \ell'}& \Sigma^{\rm TEEE}_{\ell \ell'}\\
        \Sigma^{\rm TTEE}_{\ell \ell'}  & \Sigma^{\rm TEEE}_{\ell \ell'}  & \Sigma^{\rm EEEE}_{\ell \ell'}
    \end{pmatrix},
\end{split}
\end{equation}
with 
\begin{equation}\label{eq:cosmic_variance_covmat}
\begin{split}
    \Sigma^{\rm TTTT}_{\ell \ell'}   &= \frac{2\delta_{\ell\ell'}}{f_{\textrm{sky}}\big(2 \ell + 1\big)} \big(C_\ell^{\rm TT}\big)^2  \\
    \Sigma^{\rm TTTE}_{\ell \ell'}   &=\frac{2\delta_{\ell\ell'}}{f_{\textrm{sky}}\big(2 \ell + 1\big)}  C_\ell^{\rm TT}C_\ell^{\rm TE}    \\
    \Sigma^{\rm TETE}_{\ell \ell'}   &= \frac{\delta_{\ell\ell'}}{f_{\textrm{sky}}\big(2 \ell + 1\big)} \Big(\big(C_\ell^{\rm TE}\big)^2+C_\ell^{TT}C_\ell^{\rm EE}\Big)    \\
      \Sigma^{\rm TEEE}_{\ell \ell'} &= \frac{2\delta_{\ell\ell'}}{f_{\textrm{sky}}\big(2 \ell + 1\big)}  C_\ell^{\rm EE}C_\ell^{\rm TE}      \\
    \Sigma^{\rm TTEE}_{\ell \ell'}   &= \frac{2\delta_{\ell\ell'}}{f_{\textrm{sky}}\big(2 \ell + 1\big)}  \big(C_\ell^{\rm TE}\big)^2    \\
     \Sigma^{\rm EEEE}_{\ell \ell'}  &= \frac{2\delta_{\ell\ell'}}{f_{\textrm{sky}}\big(2 \ell + 1\big)}  \big(C_\ell^{\rm EE}\big)^2 , 
\end{split}
\end{equation}
where $f_{\textrm{sky}}$ is fraction of sky observed. In this analysis, we set $f_{\textrm{sky}} = 1$ corresponding to a full sky measurement. 

In all cases, we use power spectra that cover a multipole range $2 \leq \ell \leq 5000$. For $\ell \leq 30$, the distribution of power spectra is non-Gaussian \cite[see e.g.][]{Bond:1998qg,erikson/etal:2008,Planck:2019nip,Prince:2021fdv} meaning a likelihood function would include additional terms beyond the $\Delta \chi^2$ value. We leave exploration of these non-Gaussian contributions to future work; however, we do find that excluding $\ell \leq 30$ data does not significantly alter our conclusions. 

In the data covariance, these spectra are evaluated at the fiducial cosmology described in Table \ref{tab:fiducial}. Another relevant quantity during training is $\Delta \chi^2_{\rm XY}$ that measures the emulator accuracy of each individual spectra
\begin{multline} \label{eq:delta_chi2XY}
   \Delta \chi^2_{\rm XY} \equiv \\
    \big(C_{\ell, {\rm CAMB}}^{\rm XY} - C_{\ell, {\rm emu}}^{\rm XY})^T (\Sigma^{\rm XYXY}\big)^{-1} \big(C_{\ell, {\rm CAMB}}^{\rm XY} - C_{\ell, {\rm emu}}^{\rm XY}\big) \, ,
\end{multline}
given that we train the temperature and polarization spectra independently.

\subsection{Sampling method}\label{sec:sample}

We compare two sampling methods, tempered Gaussian and uniform, for generating the training and testing sets. The parameter spaces sampled by these two methods are shown in the top panels of Figures \ref{fig:range} and \ref{fig:unirange}, respectively. Gaussian sampling naturally incorporates correlations between parameters constrained by the data, thereby reducing the emulated parameter space volume in the orthogonal directions of the CMB degeneracies. To generate the Gaussian samples, we begin by Fisher forecasting a cosmic variance limited experiment in the multipole range $30 \leq \ell \leq 400$ (using TT, TE, and EE power spectra). Any parameter correlation greater than $0.4$ is manually reduced by a factor of $2$ to help ensure proper training coverage, as the directions of CMB-induced correlations in the cosmological parameters vary as a function of the maximum adopted multipole. The overall covariance is then evenly extended by a multiplicative factor called the temperature, $T$. 

For generating the training and testing sets, we use $T_{\rm train}=256$ and $T_{\rm test}=128$, respectively. We additionally place hard prior limits, which are listed in Table~\ref{tab:parameter_ranges} for each of the cosmological parameters. For $\Omega_bh^2$ and $\Omega_ch^2$, the hard prior ranges are the same for both the training and testing sets. These hard boundaries are well beyond the range of values in the $T_{\rm train}=256$ training set. We find no evidence of hard boundary edge effects in the training of these parameters. 

Following the methodology described in Ref.~\cite{Zhong:2024xuk}, we construct a uniform sampling training set by initially generating a Latin hypercube comprising \(10^{5}\) points to efficiently cover the parameter space within the specified prior boundaries defined in Table \ref{tab:parameter_ranges} \cite{Devon/etal:2022}. Subsequently, these points are augmented by uniformly sampling additional cosmologies from the parameter space. This strategy guarantees comprehensive initial coverage while enabling an increase in the training set's size without requiring the recomputation of the entire training sample from scratch\footnote{Note that it was also found that merging multiple Latin hypercubes provides worse results \cite{Zhong:2024xuk}}. 

In most cases, the prior range in the testing set is reduced by $5\%$ to mitigate boundary effects caused by the rigid limits of the uniform sampling. However, this is not the case for $A_s$, where we apply a much more stringent cut of $\log(10^{10}A_s) < 3.5$ in the testing set. This cut is motivated in part by numerical instabilities we find in the range where $\log(10^{10}A_s) > 3.5$ in \textsc{CAMB} outputs when using \textsc{AccuracyBoost = 1.5}. We explore these numerical instabilities in more depth in Appendix~\ref{sec:cambset}.

\subsection{Pre-processing}\label{sec:train_strat_cmb}

One challenge when emulating the CMB power spectra to small scales arises from the fact that the TT and EE power spectra span several orders of magnitude. Excluding $\ell \leq 30$ polarization data, the overall amplitude of the CMB anisotropy power spectra is approximately proportional to $A_se^{-2\tau}$ \cite{Planck:2018vyg}. Consequently, we can reduce the mapping between cosmological parameters and power spectra that the emulator must learn by dividing all vectors by $A_se^{-2\tau}$; accordingly, the data vector predicted by the emulator must then be multiplied by the same factor to recover the original CMB TT, TE, and EE spectra. In Appendix \ref{app:rescale}, we model the damping tail of the lensed spectra using Genetic Algorithms to further reduce the data vector dynamical range.

A second pre-processing technique involves decomposing the variance of the CMB power spectra within the training set via a Principal Components Analysis (PCA) transformation. The PCA procedure reduces the dimensionality of the output, thereby facilitating the model training. Here, we assume the training sample contains $N_{\rm train}$ cosmologies $\boldsymbol{\theta}_k$. We compute the covariance matrix of the training set as
\begin{equation}
    \mathds{C}^{\rm XY}_{\ell_i \ell_j}=\sum_{k=1}^{N_{\rm{train}}}\frac{(C^{\rm XY}_{\ell_i}(\boldsymbol{\theta}_k)-\langle C^{\rm XY}_{\ell_i}\rangle)(C^{\rm XY}_{\ell_j}(\boldsymbol{\theta}_k)-\langle C^{\rm XY}_{\ell_j}\rangle)}{(N_{\rm{train}}-1) \sigma^{\rm XY}_{\ell_i}\sigma^{\rm XY}_{\ell_k}}\, ,
\end{equation}
where $\langle C^{\rm XY}_{\ell}\rangle$ is the average and $\sigma^{\rm XY}_{\ell}$ is the variance of the power spectra over the entire training set. 

We then construct the principal components by computing and ranking the eigenvectors $\boldsymbol K^{\rm XY}$ of the covariance  $\mathds{C}^{\rm XY}$. The dimensionality of the data vector the emulator needs to model can then be reduced from $\ell_{\rm max}$ to $N_{\rm PCA}$ by only selecting the modes that correspond to the $N_{\rm PCA}$ largest eigenvalues, as shown below
\begin{equation}
    C^{\rm XY}_{\ell}(\Vec{\theta})=\langle C^{\rm XY}_{\ell}\rangle +  \sum_{i=1}^{N_{\rm PCA}}\alpha_{(i)}(\boldsymbol{\theta}) \sigma^{\rm XY}_{\ell} K_{(i), \ell}^{\rm XY}\, ,
\end{equation}
where $\alpha_i$ corresponds to the PCA amplitudes.

\subsection{Loss Function}\label{sec:loss}

We evaluate five candidate loss functions that differ along two critical dimensions: the weighting assigned to outliers and the rescaling of the data vectors. During training, they are all evaluated as a function of $\Delta\chi^2_{\rm XY}$ as each power spectrum is trained with an independent emulator. During testing, however, our metrics, $\langle \Delta\chi^2 \rangle_{\rm med}$ and \texthtbardotlessj$(\Delta\chi^2 > 0.2)$ (outlier fraction), for evaluating the final joint emulator performance act on the data vector that concatenates all three CMB spectra and whose data covariance accounts for cross-correlations between the spectra. The first loss function candidate is
\begin{equation}\label{eq:L1}
    \mathcal{L}_1(\Delta\chi^2_{\rm XY}) =\large\langle \Delta\chi^2_{\rm XY} \large\rangle \, ,
\end{equation}
which scales quadratically in the limit $\Delta\chi^2_{\rm XY} \gg 1$. In contrast, both
\begin{equation}\label{eq:L2}
    \mathcal{L}_2(\Delta\chi^2_{\rm XY}) =\Big\langle\Big(\Delta\chi^2_{\rm XY}\Big)^{1/2}\Big\rangle \,,
\end{equation}
and 
\begin{equation}\label{eq:L3}
    \mathcal{L}_3(\Delta\chi^2_{\rm XY}) =\Big\langle\Big(1+2\Delta\chi^2_{\rm XY}\Big)^{1/2}\Big\rangle.
\end{equation}
weight outliers linearly. 

The loss $\mathcal{L}_1$ was adopted in Part I and Part II and provided outstanding results when applied to an optical weak lensing data vector. However, $\mathcal{L}_1$  overemphasizes outlier points to the extent that the model sacrifices accuracy in a large number of training models in an effort to better mitigate them. While both $\mathcal{L}_2$  and $\mathcal{L}_3$ exhibit similar behavior in handling outliers, they differ in the opposite limit, with $\mathcal{L}_3$ having the potential advantage of allowing the training to disregard further optimizations on cosmologies that are already approximated below the desired threshold $\Delta\chi^2_{\rm XY} \lesssim 0.2$.

The remaining proposed loss functions differ from $\mathcal{L}_{1,2,3}$ by being computed directly from the rescaled $\tilde{C}_\ell^{\rm XY} \equiv C_\ell^{\rm XY} \large / A_se^{-2\tau}$ instead of the original spectra. Although the emulator outputs $\tilde{C}_\ell^{\rm XY}$, we undo the rescaling prior to compute $\Delta\chi^2_{\rm XY}$ and $\Delta\chi^2$ as they depend on the original $ C_\ell^{\rm XY}$. If the data covariance rescaling varied inside the likelihood as a function of the cosmological parameters, such scaling would be irrelevant as it would leave $\Delta \chi^2$ invariant. However, this is not the case as $\Sigma$ is computed only at a fiducial cosmology. We then define both the rescaled $\Delta \tilde{\chi}_{\rm XY}^2$ that act on $\tilde{C}_\ell^{\rm XY}$ and on the rescaled covariance matrix
\begin{equation}
\Tilde{\Sigma}_{\rm XYXY} = \Sigma_{\rm XYXY} /(A_s^{\rm fid}e^{-2\tau^{\rm fid}})^2\, .
\end{equation}
The two remaining loss functions are then defined as $\mathcal{L}_{4} \equiv \mathcal{L}_{2}(\Delta \tilde{\chi}_{\rm XY}^2)$ and $\mathcal{L}_{5} \equiv  \mathcal{L}_{3}(\Delta \tilde{\chi}_{\rm XY}^2)$. 

\subsection{Training Hyperparameters}

The \textsc{learning-rate} controls the extent to which additional information influences the model's trainable parameters. In our setup, the initial \textsc{learning-rate} is set to $10^{-3}$. A learning rate scheduler then reduces this rate by a factor of two whenever the validation loss plateaus for $15$ epochs (this tunable number of epochs is the so-called \textsc{Patience}), until the \textsc{learning-rate} reaches approximately $10^{-8}$. At this point, no further reductions are applied. By progressively decreasing the learning rate, the algorithm avoids trapping the model in local minima during the early training stages while allowing minor parameter adjustments at later epochs. 

Neural network training adjusts the model parameters by comparing the model predictions against smaller batches of data vectors. The size of these batches can affect the time that the training takes to achieve convergence and the model's overall performance. However, limitations on the available RAM in GPUs restricted our choices. We fixed our batch size to $512$, which resulted in a training duration of around ten hours on an NVIDIA RTX 3060 GPU. If we set our batch size to $256$, the time for training to finish will be twice as long as our current setting, because the model will be trained on twice as many batches. We tested using a batch size of $1024$ with the same transformer model with the same amount of training data vectors, and the $\langle\Delta\chi^2\rangle_{\rm med}$ increases by a factor of $4$ and \texthtbardotlessj($\Delta\chi^2>0.2$) increases by a factor of $1.5$.

A challenge that neural networks can encounter is overfitting, in which the model becomes overly specialized to the training data and fails to generalize to unseen data. One approach to reducing the probability of such a training failure involves the introduction of penalties in the loss function that are proportional to the numerical amplitudes of the model weights. Another similar mitigation strategy forces the decay of such amplitudes. Numerically, this penalty is controlled with a parameter called the \textsc{Weight-decay}. However,  we do not run training with nonzero weight decay in this work, because our baseline model, defined in Section~\ref{sec:results}, includes trainable parameters in the activation functions that would be negatively influenced by a nonzero weight decay. Instead, we check for overfitting by visually inspecting loss values on a validation data set.

\begin{table}[t]
     \centering
     \renewcommand{\arraystretch}{1.3}
     \begin{tabular}{l|cc}\hline
          TRF          & $N_{\rm train}=200$k & $N_{\rm train}=530$k \\\hline
          Not Rescaled & $4.30$ $(1.00)$ & $0.05$ $(0.16)$ \\
          Rescaled     & $0.04$ $(0.14)$ & $0.02$ $(0.07)$\\\hline
     \end{tabular}
     
     \caption{Comparison between the same baseline transformer architectures trained with and without the rescaled power spectra. The number outside the parenthesis is the $\langle\Delta\chi^2\rangle_{\rm med}$ and the number inside the parenthesis is the fraction of outlier \texthtbardotlessj($\Delta\chi^2>0.2$).}
     \label{tab:rescale}
\end{table}

\begin{figure}[t]
    \centering
    \includegraphics[width=\columnwidth]{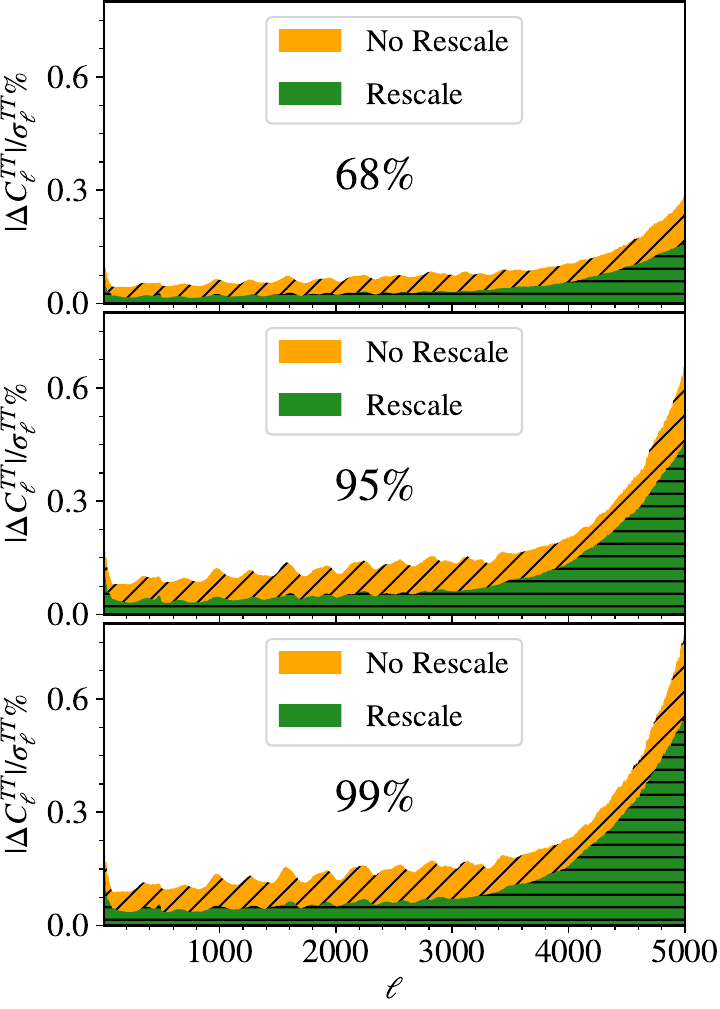}
    \caption{Distributions of the ratio between absolute error of the baseline transformer-based emulator with \textsc{CAMB} and the cosmic variance standard deviation in percentage, for the TT spectrum, with and without our rescaling scheme. The models are trained on a $T_{\rm train}=256$ Gaussian and tested on a $T_{\rm test}=128$ Gaussian set. The emulator trained on the rescaled power spectra performs better than the one trained on the non-rescaled power spectra at every $\ell$. This overall improvement results in a reduction in the value of \texthtbardotlessj($\Delta\chi^2>0.2$) by a factor of $2$. In both cases, the emulation error increases as $\ell$ increases, which we attribute to modeling the effects of the lensed CMB damping scale. In Appendix~\ref{app:rescale}, we test how including further pre-processing of the CMB damping scale can provide further improvement. }
    \label{fig:Rescalecomp}
\end{figure}

\section{Results}\label{sec:results}

In this section, we present results that quantify the performance of various emulator architectures, activation functions, loss functions, and methods for sampling parameter values for the training and testing sets. Our baseline emulator is shown schematically in Figure~\ref{fig:TRF}. It includes $N_{\rm TRF} = 1$ transformer block, and within the transformer block, the data vector is subdivided into $N_{\textrm{channel}} = 16$ individual channels to compute the self-attention, which is implemented via the dot product attention defined in Equation~\ref{attention_eq}. We fix the internal dimension of each transformer layer to be $\textsc{Int-dim-TRF}=5120$, which is approximately the same size as our data vector\footnote{Recall that we emulate the TT, TE, and EE power spectra individually, so $\textsc{Int-dim-TRF}=5120$ is approximately the same length as one of these power spectra with multipole range $2 \leq \ell \leq 5000$.} and is divisible by powers of two.

Our baseline architecture precedes the transformer block with $N_{\rm ResMLP}=3$ ResMLP blocks, each with an internal dimension of \textsc{Int-dim-ResMLP} $= 512$. Further, we adopt the H(x) function defined in Equation \ref{eq:activation_fcn} as the baseline activation function, which is one of the primary differences in our emulator compared to the emulator design presented in Part II. Unless otherwise specified, we train our emulators with parameters sampled from a tempered Gaussian distribution with temperature $T_{\textrm{train}} = 256$, and we test them with points sampled at $T_{\textrm{test}} = 128$. Finally, we employ the $\mathcal{L}_4$ loss function, which acts on the rescaled $\Delta \tilde{\chi}_{\rm XY}^2$ (see Section~\ref{sec:loss} for details on how the $\mathcal{L}_4$ loss function and the rescaled $\Delta \tilde{\chi}_{\rm XY}^2$ are defined).

We trained the baseline emulator using $N_{\rm train} = 5.3 \times 10^5$ training points. As illustrated in Figure \ref{fig:range}, our emulator replicates \textsc{CAMB} outputs. Quantitatively, the fractional difference between the baseline emulator and \textsc{CAMB} remains within cosmic variance limits for the TT, TE, and EE power spectra. This result is consistent with expectations, given that the emulator was trained against a loss function that is essentially the square root of the rescaled cosmic variance likelihood. We repeated this test using a uniformly sampled parameter space instead of a tempered Gaussian. To train this emulator, we used 1.2 million training points. We summarize the performance of the emulator in Figure~\ref{fig:unirange}, which shows that our conclusions are robust to the choice of sampling method. The runtime difference against CAMB is massive, with the Boltzmann code taking over a minute and $9$ OpenMP cores, while the emulator runtime is only $\mathcal{O}(0.01)$--$\mathcal{O}(0.1)$ seconds with a single core. In Table~\ref{tab:generalresult}, we summarize the major comparisons between a ResMLP+PCA model and a Transformer model in terms of precision, run time, and training time. The Transformer model shows higher emulation precision, while the ResMLP model requires less time to train and is faster to run during analysis. However, the difference in time consumption is not as significant as the difference in precision.

\begin{table}[t]
     \centering
     \renewcommand{\arraystretch}{1.3}
     
     \begin{tabular}{l|ccc}\hline
          Architecture & $T_{\rm train}=256$ & Run Time(s) & Train Time(hrs) \\\hline
          ResMLP+PCA    & $0.05$ $(0.19)$  & 0.08 & 8 \\
          TRF & $0.006$ $(0.06)$ & 0.30 & 18 \\\hline
     \end{tabular}
     \caption{A general comparison between the ResMLP+PCA architecture and Transformer architecture trained on $T_{\rm train}=256$ Gaussian set with $530$k points and tested on $T_{\rm test}=128$ Gaussian set. On the first column, the number outside of the parenthesis is the $\langle\Delta\chi^2\rangle_{\rm med}$ and the number inside the parenthesis is the fraction of outlier \texthtbardotlessj($\Delta\chi^2>0.2$). The second column demonstrates the approximate run time of the emulator on one cosmology on one CPU core when running in analysis. The third column shows the approximate time needed for training for each architecture under $530$k training points on an NVIDIA 3060 GPU. The ResMLP model only uses half of the training time of a Transformer model and a third of the run time of the Transformer. However, the difference in precision is significant, as the Transformer is able to achieve \texthtbardotlessj($\Delta\chi^2>0.2$)$<10\%$ with around $500$k training points, while the ResMLP model has around $20\%$ of outliers.}
     \label{tab:generalresult}
\end{table}

\subsection{Effect of Pre-Processing}

We test how pre-processing via rescaling the CMB power spectra by $A_se^{-2\tau}$, which was discussed in section \ref{sec:train_strat_cmb}, affects the emulator accuracy at fixed architecture, i.e., we do not independently re-optimize the emulator hyperparameters when comparing the loss with and without the pre-processing rescaling. In Table~\ref{tab:rescale}, we show that for a training set with $N_{\rm train} = 2 \times 10^5$ points drawn from a tempered Gaussian distribution, the data vector rescaling reduces emulators errors to the level of $\langle \Delta \chi^2 \rangle_{\textrm{med}} = 0.04$ and \texthtbardotlessj$(\Delta \chi^2 > 0.2) = 0.14$, representing an approximately order of magnitude improvement over the training without rescaling. 

With enough cosmologies in the training set, the neural network eventually achieves similar accuracy without pre-processing. This is evident by the second column in Table~\ref{tab:rescale}, where for $5.3 \times 10^5$ training points, both the rescaled and non-rescaled cases show similar performance. However, only with rescaling do we reach the desired threshold with $N_{\rm train} \lesssim 5.0 \times 10^5$. In appendix~\ref{app:rescale}, we explore an additional pre-processing of the CMB power spectra with a more sophisticated rescaling of the CMB damping tail. In particular, we based our damping tail rescaling on the analytic formulas developed in \cite{Hu:1996mn}. Lensing transfers power to small scales, resulting in a transition from pure exponential damping with the diffusion scale defined in equation \ref{eqn:damping_diff_scale} to a power-law regime. We employed a Genetic Algorithm to determine the optimal cosmological dependence of the improved scaling defined in equation \ref{eq:SR_exponent}, valid in a small region in parameter space around a fiducial cosmology.

For the case where we use only a ResMLP architecture (i.e., where we do not include a self-attention layer), we employ PCA to reduce the output dimension. This step decomposes the data into the most relevant pieces, which can reduce the number of trainable parameters in the model and the number of features that need to be learned, as the emulator neglects the least relevant principal components. This helps control RAM usage and reduces training difficulty with a smaller model that requires less training time and is easier to converge. We find that keeping the first $96$ principal components is sufficient for the emulator result to obtain $\langle\Delta\chi^2\rangle_{\rm med}\sim\mathcal
O(0.1)$, and that our results are robust to including more principal components. Higher-order principal components are mostly noisy, so they do not help generate better ML models \cite{cosmopower, Jense:2024llt}.

We show the results of including or not including the PCA pre-processing in Table~\ref{tab:pcatrf} and Table~\ref{tab:pcares} for the transformer and ResMLP architectures, respectively. We find that including the PCA pre-processing step causes the transformer-based model to lose accuracy relative to the transformer model that does not include a PCA decomposition, so we do not include the PCA pre-processing for the baseline transformer model. On the other hand, we find that the ResMLP accuracy improves by almost three orders of magnitude when using the PCA, so we do include this step when using a ResMLP-only-based architecture. We attribute the worse performance by the transformer architecture to the fact that PCA removes features that aid the attention mechanism in finding correlations in the data. 
\begin{table}[t]
     \centering
     \renewcommand{\arraystretch}{1.3}
     \begin{tabular}{l|cc}\hline
          TRF & $N_{\rm train}=200$k & $N_{\rm train}=530$k \\\hline
          PCA    & $0.550$ $(0.77)$ & $0.031$ $(0.13)$ \\
          No PCA & $0.040$ $(0.17)$ & $0.006$ $(0.06)$ \\\hline
     \end{tabular}
     \caption{Comparison between the baseline transformer architecture trained with and without first performing a PCA on the CMB power spectra. The number outside of the parenthesis is the $\langle\Delta\chi^2\rangle_{\rm med}$ and the number inside the parenthesis is the fraction of outlier \texthtbardotlessj($\Delta\chi^2>0.2$).}
     \label{tab:pcatrf}
\end{table}

\begin{table}[t]
     \centering
     \renewcommand{\arraystretch}{1.3}
     \begin{tabular}{l|cc}\hline
          ResMLP & $N_{\rm train}=200$k & $N_{\rm train}=530$k \\\hline
          PCA    & $0.10$ $(0.38)$  & $0.05$ $(0.19)$ \\
          No PCA & $75.20$ $(1.00)$ & $75.38$ $(1.00)$ \\\hline
     \end{tabular}
     \caption{Comparison between the same ResMLP architecture trained with and without PCA. The models are trained on $T_{\rm train}=256$ Gaussian set and tested on $T_{\rm test}=128$ Gaussian set. The number outside of the parenthesis is the $\langle\Delta\chi^2\rangle_{\rm med}$ and the number inside the parenthesis is the fraction of outlier \texthtbardotlessj($\Delta\chi^2>0.2$).}
     \label{tab:pcares}
\end{table}

\subsection{Transformer vs. ResMLP}

\begin{figure}[t]
    \centering
    \includegraphics[width=\columnwidth]{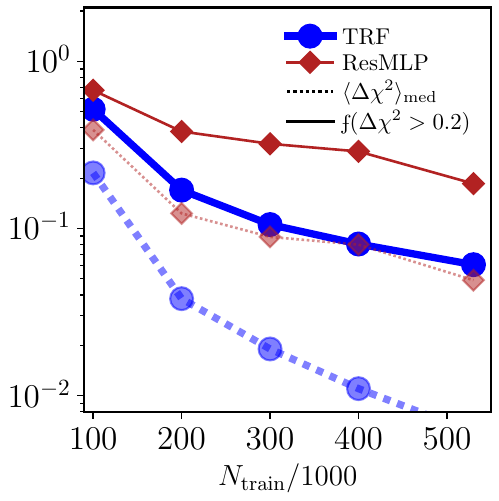}
    \caption{The median $\Delta\chi^2$ (dashed lines) and fraction of testing points with $\Delta\chi^2>0.2$, \texthtbardotlessj$(\Delta \chi^2 > 0.2)$, (solid lines) for different architectures. Thick blue lines are for our baseline Transformer ($3$ ResBlocks appended with $1$ Transformer Block), and thin red lines are for the baseline ResMLP with PCA pre-processing ($4$ ResBlocks). Each model was trained using a tempered Gaussian sampling set with $T_{\rm train}=256$ and tested with a Gaussian sampling set with $T_{\rm test}=256$. Adding additional training points results in a reduction in both accuracy metrics for both the transformer-based and ResMLP-only architectures; however, the transformer-based architecture has lower $\langle \Delta \chi^2 \rangle_{\textrm{med}}$ and \texthtbardotlessj$(\Delta \chi^2 > 0.2)$ values for every number of training points tested. While the transformer-based architecture can reduce the  \texthtbardotlessj$(\Delta \chi^2 > 0.2) < 0.1$ with around 400 thousand training points, at even 530 thousand, the ResMLP-only based architecture cannot achieve this same threshold. }
    \label{fig:resvstrf}
\end{figure}

\begin{figure}[t]
    \centering
    \includegraphics[width=0.9\columnwidth]{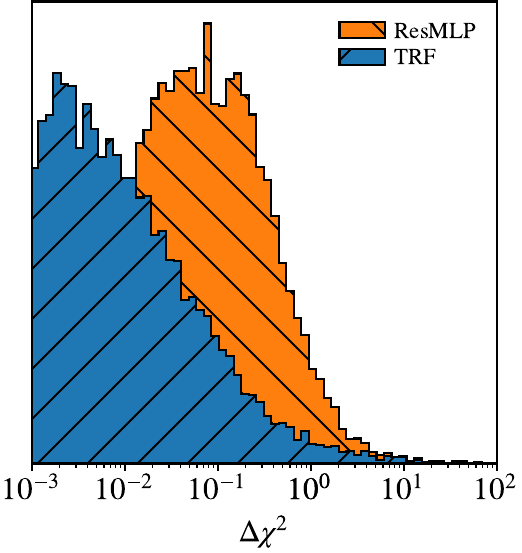}
    \caption{Comparison between the $\Delta\chi^2$ distribution of the ResMLP and transformer models trained with $530$ thousand data vectors drawn from a Gaussian sample with temperature $T_{\rm train}=256$, and then tested on a Gaussian sampling set with temperature $T_{\rm test}=128$. 
    Overall, the transformer-based emulator shifts the $\Delta \chi^2$ values toward smaller values relative to the ResMLP-based architecture. Because the entire distribution shifts toward smaller values, the median value also shifts to smaller values, and the fraction of points with $\Delta\chi^2 > 0.2$ also decreases. While we choose to use \texthtbardotlessj$(\Delta \chi^2 > 0.2)$ as our accuracy metric for outlier points, this plot illustrates that the transformer also has a lower fraction of outlier points than the ResMLP case up to $\Delta \chi^2 \approx 4$. }
    \label{restrfhist}
\end{figure}

In this subsection, we compare emulator performance between the baseline transformer architecture and a ResMLP architecture with PCA pre-processing. The ResMLP-only architecture has four ResMLP blocks, each with \textbf{Int-dim-ResMLP}$=512$. We vary the number of sample points in the training set and compare the resulting $\langle \Delta \chi^2 \rangle_{\textrm{med}}$ and \texthtbardotlessj$(\Delta \chi^2 > 0.2)$ values for both the baseline transformer and ResMLP-only architectures and show the results in Figures~\ref{fig:resvstrf} and \ref{restrfhist}. 

Increasing the number of training points generally improves the performance of both emulator architectures; however, it comes at the cost of having to compute more data vectors. For every number of training points in the sample, the baseline transformer architecture performs better than the ResMLP-only architecture for both the $\langle \Delta \chi^2 \rangle_{\textrm{med}}$ and \texthtbardotlessj$(\Delta \chi^2 > 0.2)$ metrics. For comparison, the emulator that includes a transformer block needs approximately $2 \times 10^5$ training points to achieve lower $\langle \Delta \chi^2 \rangle_{\textrm{med}}$ and \texthtbardotlessj$(\Delta \chi^2 > 0.2)$ values than the ResMLP-only based emulator achieves with $5 \times 10^5$ training points. 

Extrapolating Figure~\ref{fig:resvstrf} suggests that the ResMLP-only architecture could potentially achieve \texthtbardotlessj$(\Delta \chi^2 > 0.2)$ less than 10$\%$, the threshold we set in this analysis, with a few million data vectors. However, including a transformer block achieves this performance threshold by roughly $4 \times 10^5$ points, albeit at the cost of adding more trainable parameters, which increases the training time. For reference, our ResMLP model takes only $25\%$ of the training time of the transformer model.

In general, we find that the largest computational bottleneck in training an emulator for CMB anisotropy power spectra, both in runtime and in use of computational resources, comes from computing the data vectors and not the training times, even for models that add transformer blocks. With our accuracy settings, $9$ Intel Sapphire Rapids CPU cores need $100$ seconds to compute one data vector using \textsc{CAMB}. This highlights a key advantage of using transformer blocks for emulating CMB anisotropy power spectra. 

Figure~\ref{restrfhist} shows the distribution of $\Delta \chi^2$ values between the \textsc{CAMB} and an emulator that either includes or excludes a self-attention layer for all of the points in the testing set. The $\Delta \chi^2$ distribution for the transformer-based architecture is shifted toward smaller $\Delta \chi^2$ values than the ResMLP-only architecture. This lowers both the median $\Delta \chi^2$ value as well as suppresses the fraction of outlier points where $\Delta \chi^2 > 0.2$ as was seen in Figure~\ref{fig:resvstrf}. Importantly Figure~\ref{restrfhist} illustrates that the full distribution has shifted toward smaller $\Delta \chi^2$ values, which highlights that the transformer-based emulator should perform better than the ResMLP-only for alternative potential accuracy metrics such as \texthtbardotlessj$(\Delta \chi^2 > 0.1)$ or \texthtbardotlessj$(\Delta \chi^2 > 1)$.

\begin{figure}[t]
    \centering
    \includegraphics[width=\columnwidth]{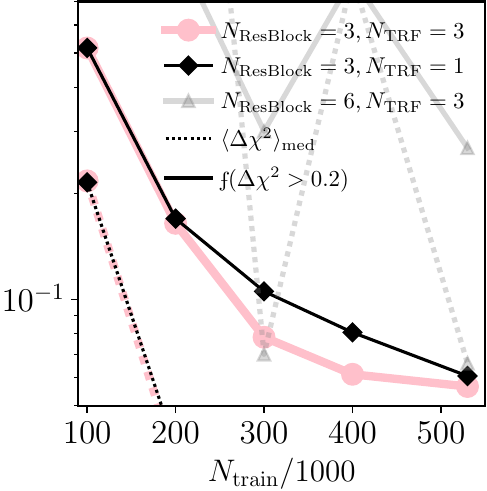}
    \caption{Comparison of the fraction of outliers, \texthtbardotlessj$(\Delta\chi^2>0.2)$, with a varying number of transformer and ResMLP blocks. The black line is for a model with one transformer block and three ResMLP blocks, the gray line is for the model with three transformer blocks and three ResBlocks blocks, and the pink line is for the model with three transformer blocks and six ResBlocks. Increasing the number of transformer blocks improves emulator accuracy at 300 and 400 thousand training vectors. However, when additional ResMLP blocks are added, the training becomes more unstable. The model does not necessarily benefit from having more data points fed during training, as we can see when the number of training data vectors increases to $400$ thousand and more. This is due to over-parametrization of the model, introducing numerical difficulties for training.} 
    \label{fig:3block}
\end{figure}

In Figure~\ref{fig:3block}, we vary the number of points in the training sample and compare the performance of the baseline emulator, which has one transformer block and three ResMLP blocks, to emulators that have three transformer blocks and either three or six ResMLP blocks. Increasing the number of transformer blocks from one to three results in an improvement in emulator accuracy for both the median and outlier points. The model with three transformer blocks has less than 10$\%$ of points with \texthtbardotlessj$(\Delta\chi^2>0.2)$ when trained with $\lesssim 3 \times 10^5$ points, whereas the model that uses only one transformer block requires  $\lesssim 4 \times 10^5$ points. However, it takes around $50\%$ more time to train on an NVIDIA 3060 GPU on the same amount of data vectors for the three transformer model, due to the larger number of trainable parameters. 

In contrast, Figure~\ref{fig:3block} also shows that increasing the number of ResMLP blocks from three to six overall worsens the performance of the emulator for both the median and outlier points, despite including more trainable parameters. We attribute this worse performance to over-parameterization of the network, which leads to training instability.

\subsection{Types of Attention Mechanisms}

\begin{figure}[t]
    \centering
    \includegraphics[width=\columnwidth]{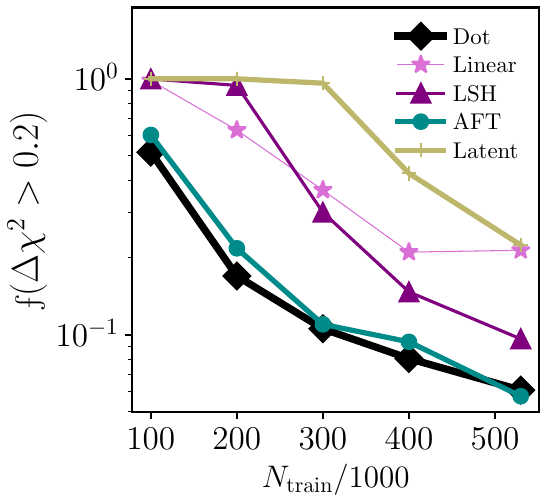}
    \caption{A comparison of the fraction of outliers, \texthtbardotlessj$(\Delta\chi^2>0.2)$, when using different attention mechanisms outlined in section~\ref{sec:att}. We find that the dot product attention and the attention-free transformer (AFT) perform about equally well. While AFT performs equally as well as the dot product attention, we find that it takes about $50\%$ more time to train the dot product attention because of the added trainable parameters. Hence, we choose to use the dot product attention mechanism in our baseline model. The other attention mechanisms, linear, latent, and LSH attention, all of which rely on an approximation, perform about $1.5-5$  times worse than the dot-product attention. }
    \label{fig:att}
\end{figure}

We have demonstrated that including a dot product attention mechanism can reduce the number of sampled training points required to achieve a set threshold accuracy for the emulator relative to ResMLP-only emulators. However, the dot product attention is not the only type of attention mechanism in the literature. Here, we compare the four alternative attention mechanisms (linear, LSH, AFT, and latent), which are discussed in more detail in Section~\ref{sec:arch} and compare them to the baseline dot product attention model\footnote{We optimize the hyperparameters for the dot-product attention mechanism and use the same values for the alternative mechanisms. We find that including these optimizations leads to a small improvement in emulator performance that does not qualitatively alter our conclusions. }. We compare the values of \texthtbardotlessj$(\Delta \chi^2 > 0.2)$ for each of the four different potential attention mechanisms as a function of the number of training points and show the results in Figure~\ref{fig:att}. Overall, we find that the baseline dot-product attention mechanism that was previously used in \cite{Zhong:2024xuk, Saraivanov:2024soy} is at least approximately as accurate for every given number of training points as each of the alternative attention mechanisms. 

The linear, latent, and LSH attention mechanisms provide alternative approaches to computing the attention matrix $QK^{T}$ in Equation~\ref{attention_eq}; they all involve approximations to reduce the time required for attention evaluation in large models. For example, the LSH attention mechanism only computes the largest products in $QK^{T}$. The latent attention, on the other hand, attempts to reduce the size of the $Q, K,$ and $T$ matrices by decomposing them into smaller matrices. 

For all five numbers of training points explored, Figure~\ref{fig:att} shows that these approximations have worse accuracy relative to the baseline dot product attention mechanisms. Quantitatively, the linear and latent attention models are roughly $4$ times less accurate for outlier points than the dot product attention models when using $5.3 \times 10^5$ training data vectors. Meanwhile, the LSH attention model is around $1.5$ times less accurate for the same number of training data vectors. Increasing the number of training points tends to result in larger improvement for the LSH method compared to the linear and dot product attention mechanisms. 

The AFT model avoids computing the attention matrix entirely, and has nearly the same accuracy level as the dot product attention model, both reaching around $6\%$ fraction of outliers when trained on $N_{\rm train}=5.3\times10^{5}$ data vectors. However, the pairwise position biases in the AFT model introduce more learnable weights compared with the dot product attention model. Since the dot product attention uses fewer computational resources while achieving the same accuracy, we consider it the optimal attention mechanism. 

We test both the positional embedding technique used in \cite{Adamo:2024blz} and the rotary positional embedding technique used in \cite{2021arXiv210409864S}. Both methods achieve the same level of precision as the dot-product attention method with $530$ thousand training data vectors and the same training procedures as above. Specifically, we find \texthtbardotlessj$(\Delta\chi^2>0.2)=0.07$ and \texthtbardotlessj$(\Delta\chi^2>0.2)=0.06$ respectively for the two positional embedding methods. We see no significant changes from applying this technique for our purpose.

\subsection{Choice of Activation Functions}
\begin{figure}[t]
    \centering
    \includegraphics[width=\columnwidth]{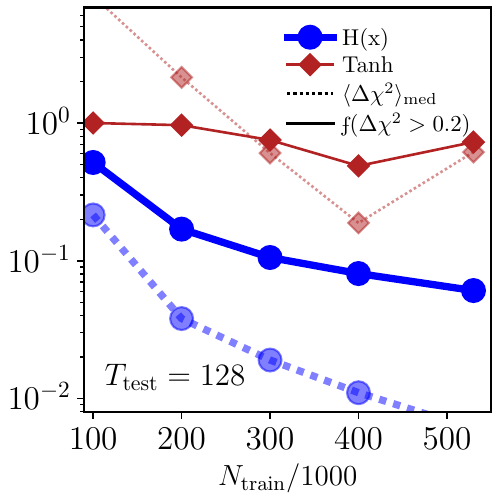}
    \caption{The median $\Delta\chi^2$ (dashed lines) and fraction of testing points with $\Delta\chi^2>0.2$, \texthtbardotlessj$(\Delta \chi^2 > 0.2)$, (solid lines) for the $\textsc{H}(x)$ (blue) and $\textsc{Tanh}(x)$ (red) activation functions. The emulators trained using the $\textsc{Tanh}(x)$ activation function perform worse than those trained with the $\textsc{H}(x)$ activation function for every number of training points tested and for both accuracy metrics. For the $\langle \Delta\chi^2 \rangle_{\rm med}$, this difference is about an order of magnitude better for the $\textsc{H}(x)$ activation function. While increasing the number of training points generally results in a lowering of the number of outlier points for the emulator using the $\textsc{H}(x)$ activation function, the emulator using the $\textsc{Tanh}(x)$ activation function does not reduce the number of outliers as the number of training points increases. These results highlight the benefit of using a dynamical activation function. }
    \label{tanh}
\end{figure}

In our baseline emulator, we use the $\textsc{H}(x)$ activation function used previously in \cite{Alsing/etal:2020,cosmopower}. In this subsection, we compare the performance of the emulator using this activation function to the $\textsc{Tanh}$ function that was used in Parts I and II. In Figure~\ref{tanh}, we show this comparison for the median $\Delta \chi^2$ value and the \texthtbardotlessj$(\Delta\chi^2>0.2)$  for varying numbers of training points. We find that the $\textsc{H}(x)$ activation function generally performs better for both accuracy metrics than $\textsc{Tanh}$. Increasing the number of training points reduces the median $\Delta \chi^2$ value for both activation functions, though there is a notable uptick in the median $\Delta \chi^2$ value when increasing from $4 \times 10^5$ to $5.3 \times 10^5$ training points. Nevertheless, the $\textsc{H}(x)$ activation function performs about an order of magnitude better at each number of training points. For the \texthtbardotlessj$(\Delta \chi^2 > 0.2)$ value, the $\textsc{Tanh}$ activation function shows negligible improvement when increasing the number of training points from $10^5$ to $5.3 \times 10^5$. Over the same range, the $\textsc{H}(x)$ activation goes from being $1.5$ times smaller than the corresponding $\textsc{Tanh}$ outlier fraction to being $10$ times smaller. 

\subsection{Choice of Loss Functions}

\begin{figure}[t]
    \centering
    \includegraphics[width=\columnwidth]{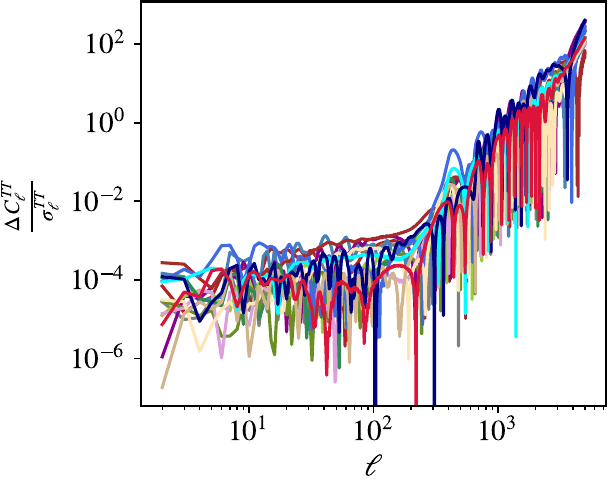}
    \caption{The difference between the TT power spectrum outputs of the baseline emulator trained with the Mean Square Error (MSE) loss function and \textsc{CAMB} divided by the square root of the diagonal of the cosmic variance covariance matrix for the $15$ cosmologies from Figure~\ref{fig:range}. This difference is shown per $\ell$ mode. As the multipole moment increases, the MSE-trained emulator results in larger errors relative to cosmic variance. We attribute this larger relative error to MSE equating differences in power on large scales, where the overall power is much larger, to differences in power on small scales. This causes the emulator to be biased towards fitting the larger scales more optimally. This motivates us to use the loss functions based on the cosmic variance $\Delta \chi^2_{\rm XY}$, which are defined in Section~\ref{sec:loss}.}  
    \label{fig:mse}
\end{figure}

An optimal choice of loss function can help improve the accuracy of the emulator for some set number of training points and reduce the amount of time necessary to train the emulator. Previous works, such as \cite{cosmopower}, have used the Mean Square Error (MSE) as the loss function for training CMB emulators. We trained our baseline emulator using the MSE loss function and $530$ thousand training points. In Figure~\ref{fig:mse}, we show the difference between the outputs of the emulator trained with the MSE loss function and \textsc{CAMB} divided by the square root of the diagonal component of the cosmic variance covariance matrix for the $15$ randomly chosen cosmologies in Figure~\ref{fig:range}. We find that the emulation error relative to cosmic variance increases as multipole increases, with the emulation error exceeding cosmic variance for $\ell \gtrapprox 2000$ and approaching $100$ times larger than cosmic variance by $\ell = 5000$. 

The baseline emulator is trained using the $\mathcal{L}_4$ loss function defined in Section~\ref{sec:loss}. One simple alternative is the $\mathcal{L}_1$ loss function of Equation~\ref{eq:L1}, which is the average $\Delta \chi^2_{\rm XY}$\footnote{Recall that we train the emulators for individual power spectra, so the subscript XY simply denotes the relevant power spectra that is being emulated (e.g. TT). We adopt this notation to differentiate the $\Delta \chi^2_{\rm XY}$ used in the loss function with the $\Delta \chi^2$ that includes all three concatenated power spectra that we use to assess the accuracy of our emulators primarily through $\langle \Delta \chi^2 \rangle_{\textrm{med}}$ and \texthtbardotlessj$(\Delta \chi^2 > 0.2)$.} value  between the emulator and \textsc{CAMB}. However, we find that this loss function performs orders of magnitude worse for both the median $\Delta \chi^2$ value and the fraction of outlier points, \texthtbardotlessj$(\Delta \chi^2 > 0.2)$, than the baseline case for every number of training points tested. 

\begin{figure*}[t]
    \centering   
    \includegraphics[width=\columnwidth]{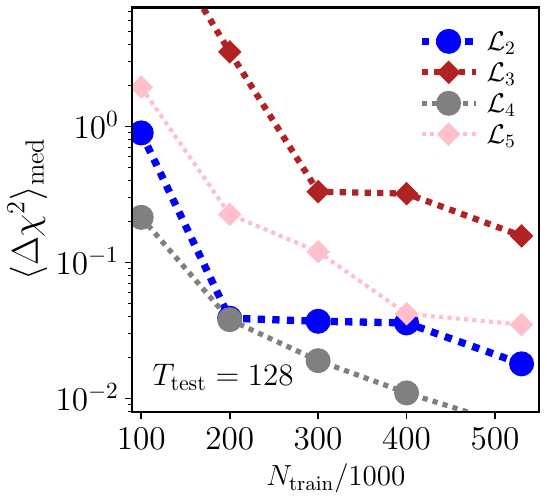}
    \includegraphics[width=\columnwidth]{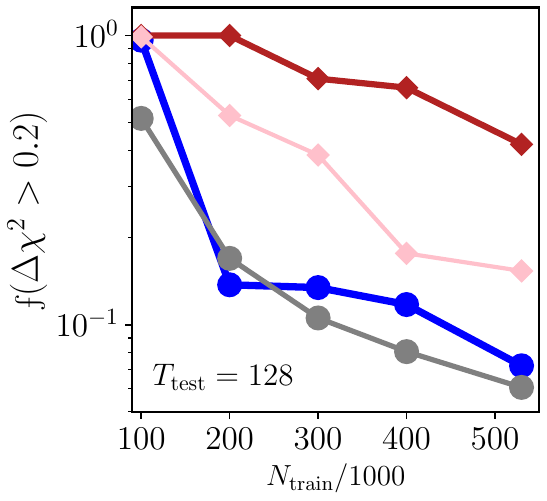}
    \caption{
    \emph{Left panel}: Comparison between the median $\Delta\chi^2$ for each loss function outlined in Section~\ref{sec:loss}. We exclude plotting $\mathcal{L}_1$ because the lowest median $\Delta\chi^2$ we observed was about $10^2$. We find that the function $\mathcal{L}_4$ scales best with the addition of more training data while also providing the smallest median $\Delta\chi^2$. The next best function, $\mathcal{L}_2$, has a median $\Delta\chi^2$ that is higher than $\mathcal{L}_4$ for every number of training points except 200 thousand. Both $\mathcal{L}_3$ and $\mathcal{L}_5$, improve with more training points; they never exceed the performance of $\mathcal{L}_4$.
    \emph{Right panel}: Comparison between the \texthtbardotlessj$(\Delta\chi^2>0.2$) for each loss functions. In agreement with the left panel, the functions $\mathcal{L}_2$ and $\mathcal{L}_4$ perform approximately equally well to each other while performing significantly better than the other loss functions. However, between $\mathcal{L}_3$ and $\mathcal{L}_5$, we find that the rescaled option $\mathcal{L}_5$ is significantly better than the non-rescaled version. From these plots, we conclude that $\mathcal{L}_4$ is the best loss function to use for our CMB emulator.}
    \label{fig:loss}
\end{figure*}

Figure ~\ref{fig:loss} shows the accuracy metrics, $\langle \Delta \chi^2 \rangle_{\textrm{med}}$ and \texthtbardotlessj$(\Delta \chi^2 > 0.2)$, for each of the $\mathcal{L}_2$, $\mathcal{L}_3$, $\mathcal{L}_4$, and $\mathcal{L}_5$ loss functions given some number of training points. In general, all of the loss functions show improvement in accuracy as the number of training points increases. However, emulators trained using the $\mathcal{L}_2$ loss function hit a plateau for both accuracy metrics between $2 \times 10^5$ and $4 \times 10^5$ training points. Using the hyperbolic loss functions, $\mathcal{L}_3$ and $\mathcal{L}_5$, results in generally less accurate emulators than using the square root of the $\Delta \chi^2_{\rm XY}$ values as is done in $\mathcal{L}_2$ and $\mathcal{L}_4$. In Appendix~\ref{sec:3x2pt}, we show results updating the optical weak lensing $3\times2$pt emulator from part II with a hyperbolic loss function similar to $\mathcal{L}_3$, where we find that this loss function reduces emulation error compared to the loss function in part II. 

The $\mathcal{L}_5$ loss function performs approximately as well as the $\mathcal{L}_2$ on the $\langle \Delta \chi^2 \rangle_{\textrm{med}}$ values, highlighting that rescaling the covariance matrix and \textsc{CAMB} outputs can help improve the accuracy of the emulator. This is further supported by the performance of the $\mathcal{L}_4$ loss function relative to $\mathcal{L}_2$. However, an emulator trained with the $\mathcal{L}_5$ loss function tends to perform worse than emulators trained with either $\mathcal{L}_2$ or $\mathcal{L}_4$ for outlier points in the tails of the distribution. This latter point is corroborated by the baseline $\mathcal{L}_4$ loss function performing as well or better than the $\mathcal{L}_2$. When trained with $5.3\times10^{5}$ data vectors, the model trained with $\mathcal{L}_4$ can eliminate outliers to $6\%$, and the one with $\mathcal{L}_2$ has $7\%$ of outliers. Notably, even without the rescaling that is unique to CMB, a similar loss function, $\mathcal{L}_2$, can achieve performance on par with $\mathcal{L}_4$. Hence, this result of $\mathcal{L}_2$ can likely be generalized to other emulation scenarios. On the other hand, $\mathcal{L}_4$ is only physically motivated for CMB. 

\subsection{Choice of Training Sampling Methods}\label{sec:sampling_results}
In previous sections, we have performed all tests using the $T_{\rm train}=256$ Gaussian sampled training set, where we can achieve our criteria of \texthtbardotlessj$(\Delta\chi^2>0.2) < 0.1$ for points in the testing set with around $4\times10^{5}$ training points. In comparison, we find that training on the uniformly sampled training set makes it more difficult to control this fraction of outliers. To explore this further, we test the same transformer architecture used in the previous tests, but trained and tested on the uniform samples (see Table~\ref{tab:parameter_ranges} for the parameter ranges). 

\begin{figure}[t]
    \centering
    \includegraphics[width=\columnwidth]{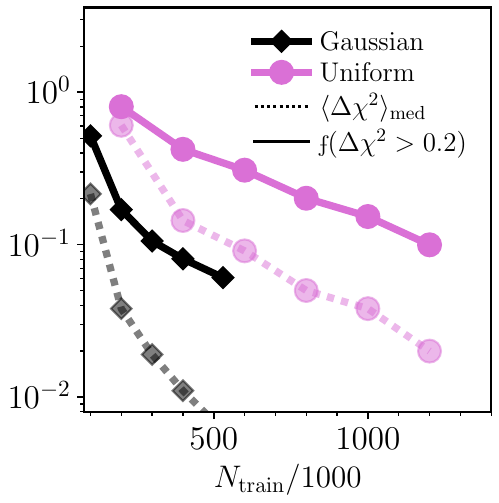}
    \caption{Comparison between the uniform and the Gaussian sampling methods. The solid lines display the \texthtbardotlessj$(\Delta\chi^2>0.2)$ and the dashed lines display the median $\Delta\chi^2$. 
    The blue lines are for models trained on Gaussian samples on $T_{\rm train}=256$ and tested on $T_{\rm test}=128$, and the red lines are trained on the uniform samples drawn from the parameter ranges outlined in Table~\ref{tab:parameter_ranges}. The Gaussian sampling provides around an order of magnitude improvement in both of the metrics displayed in the plot. However, with enough training data, about three to five times more, the uniform sampling can perform approximately as well as the Gaussian sampling. }
    \label{uniformvsgauss}
\end{figure}

As we can see in Figure~\ref{uniformvsgauss}, even with $6\times10^{5}$ training data vectors, the uniform sampling trained model still has \texthtbardotlessj($\Delta\chi^2>0.2$)$ \ = 31\%$. Nevertheless, there is a downward trend in the accuracy metrics of the emulator trained and tested on the uniform sets. Thus, we further apply up to $1.2\times10^{6}$ uniformly sampled training data vectors, where we are able to achieve \texthtbardotlessj($\Delta\chi^2>0.2$)$=10\%$ on the uniform testing set. This shows that, given enough training data, it is possible to train on parameters uniformly sampled from the prior, although it requires a tighter cut on $\log(10^{10}A_s)$ than the Gaussian case. 

\subsection{Interpolation Ability} \label{Interpolation}

In our baseline emulator, the models are trained on a Gaussian training set using a temperature $T_{\rm test}=T_{\rm train}/2$. However, when expanding a $d$-dimensional Gaussian distribution by tempering the covariance this way, the density of points generated near the fiducial cosmology decreases approximately as $T^{-d}$ because the same number of points now covers a larger parameter space. This could mean that a model trained over a large temperature parameter distribution has not seen enough data near the fiducial cosmology to accurately model points sampled using a much lower temperature. 

If the high-temperature testing set also has too few points near the fiducial model, then our metrics might miss the poor performance of the emulator near the fiducial cosmology. This is particularly relevant as the emulator would likely be used for future CMB data whose allowed parameter space will be smaller than even the parameter space covered by our $T=1$ tempered Gaussian. To verify that our models are not sensitive to this effect, the model trained on $T_{\rm train}=256$ is tested on samples generated from Gaussian distributions with decreasing temperatures, down to $T=1$. 

\begin{figure}[!ht]
    \centering
    \includegraphics[width=0.95\columnwidth]{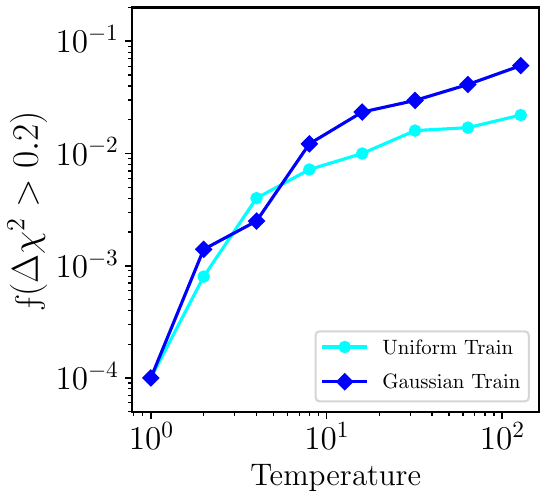}
    \caption{The fraction of outliers, \texthtbardotlessj$(\Delta\chi^2>0.2)$ as a function of the testing temperature, $T_{\rm test}$. We train a transformer model on a Gaussian training set with $T_{\rm train}=256$ and another model with uniform sampling, and test the model on all temperatures in powers of two between $T_{\rm test}=128$ and $T_{\rm test}=1$. To test the models trained on the uniform sets, we restrict $\log{(10^{10}A_s)}\leq3.5$. We find, in both cases, that the \texthtbardotlessj$(\Delta\chi^2>0.2)$ tends to zero as the temperature decreases to one, suggesting that, as the training temperature increases and the density of training points near the fiducial cosmology decreases, our emulators are still accurately able to replicate \textsc{CAMB} outputs near the fiducial cosmology. This is important because current and future CMB experiments will allow only a small volume parameter around the fiducial cosmology. }
    \label{fig:temp}
\end{figure}

Despite the decrease in the density of training points, we can see from Figure~\ref{fig:temp} that the fraction of outliers tends to $0$ as we decrease the temperature of the testing set. We additionally tested the performance of the emulator on a Planck $\Lambda$CDM sample chain using baseline PLIK half mission likelihoods over $30\leq\ell\leq2508$ in TT and $30\leq\ell\leq1996$ for TE and EE, HFI low-$\ell$ ($2<\ell<29$), and lensing data. Note that this parameter space is slightly smaller than our $T_{\rm test}=1$ because our cosmic variance-limited Fisher matrix that we use to generate sample points only includes multipoles $30 \leq \ell \leq 400$ for the TT, TE, and EE power spectrum. When using this Planck sample chain, we find no outliers (i.e., testing points with $\Delta \chi^2$ values greater than $0.2$).

Furthermore, we show that a model trained with $1.2$ million uniform sampled data vectors also has good interpolation ability in small regions around the fiducial point. Due to the different prior cuts we have for training in uniform sampling and Gaussian sampling, we test this model with a parameter cut of $\log{(10^{10}A_s)}\leq3.5$. We see from Figure~\ref{fig:temp} that the fraction of outliers from this model also decreases to $0$ as we reduce the temperature of the Gaussian testing set. From this, we can see that either method of generating training samples can be used to run an MCMC. However, as seen in section~\ref{sec:sampling_results}, the Gaussian set simply achieves better accuracy with fewer training points. With enough training data, the model trained on uniform samples could achieve the same accuracy as the Gaussian trained samples at all temperatures, but with a smaller prior in $\log{(10^{10}A_s)}$.

\section{Conclusion} \label{sec:conclusion}

In this work, we emulate CMB TT, TE, and EE power spectra calculated by the \textsc{CAMB} Boltzmann code for the $\Lambda$CDM model. We extend previous analyses such as Ref.~\cite{cosmopower} that also emulated CMB anisotropy power spectra outputs of Boltzmann codes in two ways: (1) we broaden, relative to these previous analyses, the $\Lambda$CDM parameter space for which the emulator is validated, and the multipole range of the power spectra emulated, and (2) we improve the emulator accuracy to be within cosmic variance uncertainties. This analysis also builds on Ref.~\cite{Zhong:2024xuk} and Ref.~\cite{Saraivanov:2024soy} that applied self-attention mechanisms to optical weak lensing data. Overall, we conclude that we can emulate the CMB anisotropy power spectra outputs of \textsc{CAMB} to within cosmic variance limits out to $\ell \approx 5000$ while covering a very large parameter space. 

We find that using a transformer architecture, which includes a self-attention mechanism, improves the emulator performance, for all numbers of training points tested, relative to ResMLP-only architectures for both points near the fiducial cosmology as well as for points in the tails of the distribution. This latter point is important because it allows the emulator to cover a broader parameter space with fewer training points. Importantly, one may be able to achieve similar performance as the transformer architecture with a ResMLP-only architecture; however, this will require on the order of hundreds of thousands more training data vectors to achieve. 

While the transformer architecture adds more trainable parameters relative to a ResMLP-only architecture, which in turn increases the time necessary to train the model, we conclude that it is still beneficial to use the transformer architecture. In general, we find the most time-consuming and computationally expensive part of emulating CMB power spectra is the initial calculation of all of the data vectors using \textsc{CAMB} to train the emulator. Therefore, reducing the number of training vectors necessary to emulate the power spectra greatly expedites cosmological analysis. 

We tested including more self-attention layers as well as changing how the self-attention in the transformer blocks is calculated. However, we found at most marginal improvements relative to the improvement found by whether the model includes a self-attention layer or not. In Appendix~\ref{sec:cnn} we test using a 1D convolutional neural network, which performs about as well as the transformer. Additionally, we found that the dynamical activation function with learnable weights adopted by by~\cite{cosmopower} considerably improves the emulation accuracy and efficiency compared to static activation functions like $\textsc{Tanh}$.

Pre-processing the CMB data vectors can significantly improve emulator accuracy as it can reduce the amount of parameter correlations that the emulator needs to learn. In particular, we found improvement when rescaling the amplitude of the CMB data vectors by $A_se^{-2\tau}$ for both the transformer and ResMLP-only architectures. In Appendix~\ref{app:rescale}, we show that rescaling the amplitude of the damping tail may also improve emulator performance. 

Optimally sampling the parameter space can improve the performance of the emulator while also allowing for good coverage of a broad parameter space. Specifically, we found, as it was previously found in Ref.~\cite{Saraivanov:2024soy}, that using a tempered Gaussian sampling method reduces the number of training data vectors necessary to achieve a set emulation accuracy threshold relative to uniform sampling. We attribute this improvement to taking into account the correlation between parameters when generating the larger parameter space so that the emulator does not need to learn to emulate cosmologies that are far from the fiducial parameters in a direction orthogonal to the CMB parameter degeneracies. In effect, we are creating a more targeted parameter volume to emulate. Increasing the volume of parameter space where the emulator is valid is beneficial as a proxy for extended cosmological models, as well as a tool for collaborations employing blinding methods to shift the data vector by an arbitrary amount and still achieve reasonable results from the emulator. 

Improving emulator accuracy and precision to be in agreement with \textsc{CAMB} to within cosmic variance is valuable because this enables the emulator to be applicable to real data for the next couple of decades of CMB analyses. In Appendix~\ref{app:exp}, we show this explicitly using the actual Planck data covariance matrix as well as forecasts for Simons Observatory-like, CMB-S4-like, and CMB-HD-like experiments. We additionally take an Atacama Cosmology Telescope (ACT) DR6 data MCMC \cite{ACT:2025tim} and calculate the shift in $\chi^2$ when replacing \textsc{CAMB} with our baseline emulator and find the \texthtbardotlessj$(\Delta\chi_{\rm ACT}^2>0.2)<9\%$ with only $200$ thousand training points on uniform sampling testing set. With the same emulator trained and tested on Gaussian sampling, \texthtbardotlessj$(\Delta\chi_{\rm ACT}^2>0.2)$ is further suppressed to under $0.1\%$. We conclude that our emulator has broad applicability to both current and future CMB experiments. 

\section*{Acknowledgments}

Special thanks to Kunhao Zhong for helpful discussions about the convolutional neural networks. VM and YZ were partially supported by the Roman Project Infrastructure Team ``Maximizing Cosmological Science with the Roman High Latitude Imaging Survey" (NASA contracts 80NM0018D0004-80NM0024F0012). V.M. was also partially supported by the Roman Project Infrastructure Team ``A Roman Project Infrastructure Team to Support Cosmological Measurements with Type Ia Supernovae" (NASA contract 80NSSC24M0023). E.S.  acknowledges support from the Yang Institute for Theoretical Physics and Stony Brook University as a grad assistant, partly made possible by generous contributions from the Simons Foundation.  The work of A.S.G. is supported in part by the National Science Foundation grant PHY-2210533. Simulations in this paper use High Performance Computing (HPC) resources supported by the University of Arizona TRIF, UITS, and RDI and maintained by the UA Research Technologies department. The authors would also like to thank Stony Brook Research Computing and Cyberinfrastructure, and the Institute for Advanced Computational Science at Stony Brook University for access to the high-performance SeaWulf computing system. 

\appendix

\section{\textsc{CAMB} Settings}\label{sec:cambset}
We adopt the \textsc{CAMB} settings in Appendix A of \cite{MacInnis:2023vif}; however, we increase the \textsc{Accuracy-Boost} parameter to 1.5 instead of 1.1. We assume results from this setting as our `truth' value for CMB power spectra, and use the same setting for training, testing, and validation sets as different settings, including $\ell_{\rm max}$, can numerically alter the power spectra. We also do not consider the comparison between numerical outputs between \textsc{CAMB} and \textsc{CLASS}\cite{2011arXiv1104.2934L}.
\begin{table}[t]
     \centering
     \renewcommand{\arraystretch}{1.3}
     \begin{tabular}{m{10em}m{3cm}}\hline
          \textsc{CAMB} setting & Value \\\hline
          \textsc{Accuracy-Boost} & $1.5$ \\
          $\ell$\textsc{-Sample-Accuracy} & $10$\\
          $\ell$\textsc{-Accuracy-Boost} & $3$\\
          \textsc{Do-Late-Rad-Truncation} & \textsc{False}\\
          $\ell_{\rm max}$ & $15000$\\
          \textsc{Lens-Potential-Accuracy} & $30$\\
          \textsc{Lens-Margin} & $2050$\\
          \textsc{camb-params-NonLinear} & \textsc{camb-model-NonLinear-both}\\
          \textsc{camb-params-NonLinearModel} & \textsc{mead2016}\\\hline
     \end{tabular}
     \caption{The \textsc{CAMB} settings we use to generate our power spectra. Our choices are based on \cite{MacInnis:2023vif}; however, we specifically changed \textsc{Accuracy-Boost} to $1.5$ for more robust accuracy. Future works may consider even higher settings for \textsc{Accuracy-Boost}.}
\end{table}
We noticed that there is some numerical inaccuracy that is visible under cosmic variance covariance in the power spectrum generated by this setting, when we compared those data vectors to some data set generated by more accurate settings. 

\begin{figure}[t]
    \centering
    \includegraphics[width=\columnwidth]{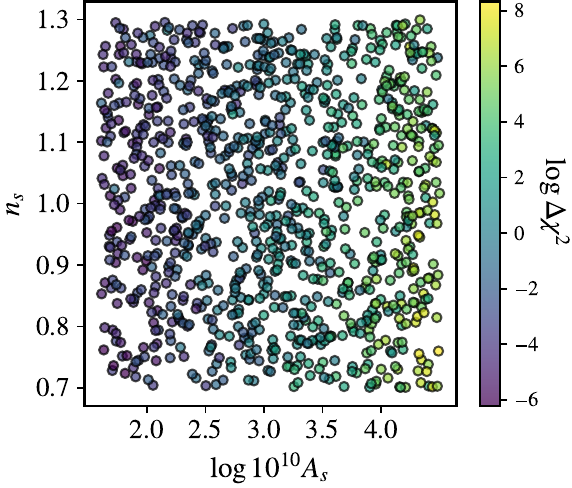}
    \caption{Distribution of $\log{(\Delta\chi^2)}$ for a cosmic variance limited experiment between \textsc{CAMB} outputs with \textsc{Accuracy-Boost}$=1.5$ and $1.8$, with all other settings fixed. There are significant deviations between outputs with those two settings in the high $A_s$ region. } 
    \label{fig:outab}
\end{figure}

In Figure~\ref{fig:outab}, we compare $\log{(\Delta\chi^2)}$ under cosmic variance covariance between \textsc{CAMB} outputs with \textsc{Accuracy-Boost}$=1.5$ and $1.8$, and noticeably, the magnitude of this inaccuracy increases as $A_s$ increases, which we attribute to higher $A_s$ values increasing the effect of lensing. Thus, for the application of our models, users need to be aware of the necessity of picking a satisfactory accuracy setting. For future works, we need to set the \textsc{Accuracy-Boost} to even higher values, possibly around $2.5$. We note for our purposes that potential \textsc{CAMB} numerical instabilities, relative to cosmic variance uncertainties, resulting from lower \textsc{Accuracy-Boost} settings would make training an emulator more difficult. This is because the emulator will attempt to learn these numerical instabilities, which are inherently random. Using a higher \textsc{Accuracy-Boost} setting could potentially lower the number of training data vectors necessary to train an emulator to within a desired accuracy. 

\begin{figure}[t]
    \centering
    \includegraphics[width=\columnwidth]{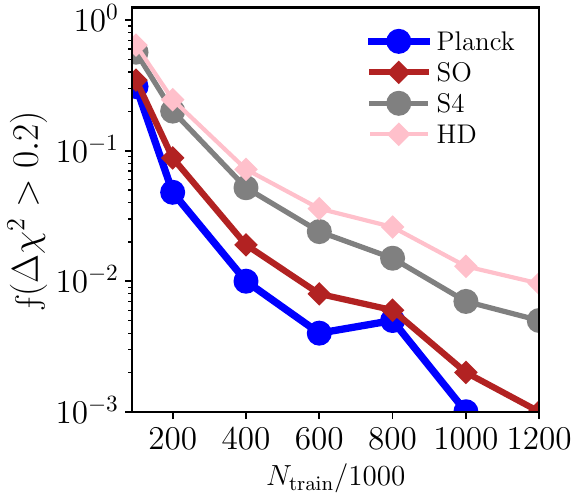}
    \caption{Comparison of the fraction of outlier points, \texthtbardotlessj$(\Delta\chi^2>0.2)$, given configurations for Planck-like (blue), Simons Observatory-like (red), CMB-S4-like (gray) and CMB-HD-like (pink) experiments. With the Planck-like and Simons Observatory-like analyses, an emulator trained on uniform sampling set with $200$ thousand training data vectors would be sufficient in precision, while for more advanced configurations in the future, like a CMB-S4-like or CMB-HD-like experiment, we will need around $400$ thousands training points to acquire the necessary precision. }
    \label{experiments}
\end{figure}

\section{Emulator Performance Estimation for Current and Future Experiments} \label{app:exp}

We test our emulator performance given Planck-like uncertainties as well as forecasted data covariances for Simons Observatory-like, CMB-S4-like, and CMB-HD-like experiments in Figure~\ref{experiments}, with binning strategies covariance matrices from \cite{Planck:2018vyg,Planck:2018nkj,Planck:2019nip, MacInnis:2023vif}. For all experiments, $\ell_{\rm min}=30$. For the Planck-like experiment, $\ell_{\rm max}=2508$ for the TT power spectrum and $\ell_{\rm max}=1996$ for TE and EE. For the TT power spectrum from the Simons Observatory-like and CMB-S4-like experiments, $\ell_{\rm max}=3000$. For the Simons Observatory-like and CMB-S4-like experiments TE and EE power spectra, $\ell_{\rm max}=5000$. For CMB-HD, the collaboration plans to measure $ 1000 \leq \ell \leq 20000$. However, our emulator was only trained to give outputs up to $\ell = 5000$, so we restrict the maximum multipole range from the CMB-HD-like forecast to $\ell = 5000$. We assume for the CMB-HD-like forecast that they can append data taken from Simons Observatory and CMB-S4 for the lower $\ell$ range, and test up to the upper limit of our emulator, so we perform the forecast for the CMB-HD-like experiment in the range of $30\leq\ell\leq5000$.

\begin{figure}[t]
    \centering
    \includegraphics[width=0.77\columnwidth]{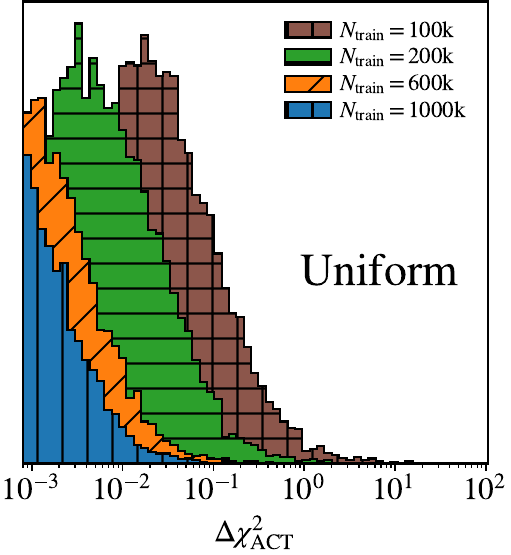}
    \includegraphics[width=0.77\columnwidth]{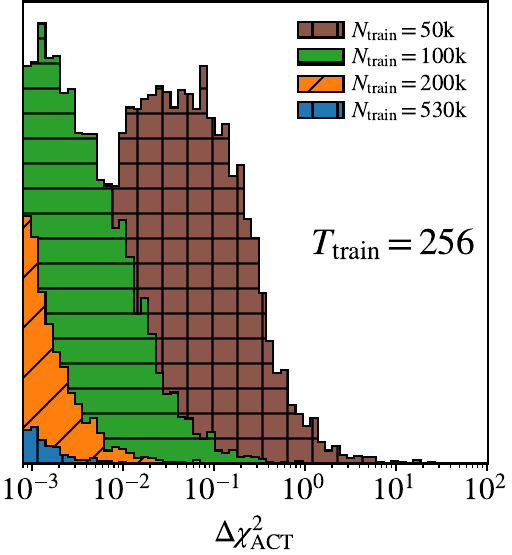}
    \caption{$\Delta\chi^2_{\rm ACT}$ of emulators trained on $100$, $200$, $400$ and $1000$ thousand data vectors on uniform sampling and the ones trained on $50$, $100$, $200$, $300$ and $530$ thousand data vectors on Gaussian sampling. All computations are evaluated with the binning and covariance matrix in ACT-DR6-CMB only log-likelihood. For the uniform sampling models, the deviations are well controlled under the level of $10^{-1}$ with $600$ thousand training data vectors, and for the Gaussian sampling models, the deviations are well controlled under the level of $10^{-1}$ with $200$ thousand training data vectors. The \texthtbardotlessj$(\Delta\chi_{\rm ACT}^2>0.2)$ is each $9\%$, $1\%$, $0.2\%$ and $0\%$ for the tests we perform on uniform sampling, and is each $16\%$, $0.4\%$ for the Gaussian sampling models with $50$ and $100$ thousand training data vectors, and no outliers for models trained with $200$ thousand and above.}
    \label{act}
\end{figure}

We find that for the Planck-like and Simons Observatory-like cases, we can train emulators on uniform sampling sets with $200$ thousand training data vectors to achieve \texthtbardotlessj$(\Delta\chi^2>0.2) \leq 10\%$. For the CMB-S4-like and CMB-HD-like forecasts, which have higher precision, we require around $400$ thousand training data vectors to train a reliable emulator. As expected, when we target realistic experiments, we do not need as many training data vectors as when we train with a cosmic variance-limited covariance matrix on every multipole. Note that we compute with ranges set by experiment limits and our current emulator training setting, described in Figure~\ref{experiments}. More detailed verification for some exact experimental configurations will be necessary in the actual application.

In the light of the new ACT data release \cite{ACT:2025tim}, we apply the new ACT CMB power spectra only likelihood from their Data Release 6 (DR6) to compare the $\Delta\chi^2_{\rm ACT}$ between \textsc{CAMB} outputs and emulator outputs under their binning and covariance matrix, up to $\ell=5000$. For the emulator, we use the same baseline transformer architecture trained with a uniform sampling set with various numbers of training points. We ignore higher multipole contributions due to the emulator training range and BB power spectrum, since we have not yet explored this signal. We show in Figure~\ref{act} that $\Delta\chi^2_{\rm ACT}$ of the emulator is well bounded below $10^{-1}$ with around $600$ thousand training data vectors on uniform sampling, and with around $200$ thousand training data vectors on Gaussian sampling, supporting that these emulators are reliable to replace \textsc{CAMB} in actual MCMC chains. 

\section{Weak Lensing and Galaxy Clustering Emulator} \label{sec:3x2pt}

The emulators presented in parts I and II~\cite{Zhong:2024xuk,Saraivanov:2024soy} each present outstanding results for emulating weak lensing observables. We test how the techniques explored in this work can improve the weak lensing emulators. To begin, we observed that simply tempering the parameter covariance matrix resulting from a forecasted LSST Y1 Fisher analysis does not completely populate the parameter space (see Figure $7$ in~\cite{Saraivanov:2024soy}). This results from a combination of hard priors and strong correlations in the sampling parameters, particularly with $\Omega_{\rm b}$. As such, we reduce the correlations of all parameters with $\Omega_{\rm b}$ by a factor of $2$, but we did not reduce the autocorrelation of $\Omega_{\rm b}$.

Next, the $3\times 2$pt data vectors require more parameters to enter the emulation and result in a larger data vector. Correspondingly, we expand the internal dimensions of the neural network as well. The input dimension of 22 (cosmological parameters, five source photo-$z$, five lens photo-$z$, two intrinsic alignment, and 5 galaxy bias) is mapped to an internal dimension of $512$ in the \textsc{ResMLP} portion of the network. The transformer portion has an internal dimension of $3840$ with $60$ channels, reflecting the number of sample bins in the resulting data vector. The output data vector has a dimension of $1560$. 

Finally, and perhaps most importantly, in an effort to minimize the effect of outliers, we adopt an outlier-friendly loss function that will prevent large errors from degrading the overall performance of the emulator. In particular, we use a hyperbolic loss function given by $\mathcal{L}(\Delta \chi^2_{\rm XY}) = \langle \sqrt{1 + \Delta \chi^2_{\rm XY}}\rangle$, which is similar to the $\mathcal{L}_3$ loss function defined in Equation~\ref{eq:L3}. This provides an enormous improvement over the results of part II, where outliers dominated the overall performance of the emulator. This change allows us to use just $10^5$ training points. With fewer training points, we correspondingly reduce the batch size to $32$.

There are a few things to note about these changes. First, the \textsc{ResTRF} tends to greatly benefit from smaller batch sizes. With $10^5$ training points and a batch size of $256$, we find the \textsc{ResMLP} slightly outperforms the \textsc{ResTRF}, demonstrating the importance of choosing hyperparameters. Both architectures, however, perform better with a smaller batch size. The larger internal and output dimensions provide a computational cost, as the size of the output layer is much larger than in the cosmic shear case. Thus, while the $3\times 2$pt emulator is slower than the cosmic shear, it is still orders of magnitude faster than numerical methods. Lastly, due to redshift space distortions being used to compute the galaxy-galaxy autocorrelation, the galaxy bias is not a fast parameter and must be included in the emulation. 

With these changes, the results are significantly improved over those in Part II. We train the emulator with temperatures in powers of two up to $T_{\rm train}=1024$, which was the largest value we set before encountering errors. A summary of the results can be found in Table~\ref{tab:3x2pt_results}. 

\begin{figure}
    \centering
    \includegraphics[width=0.9\columnwidth]{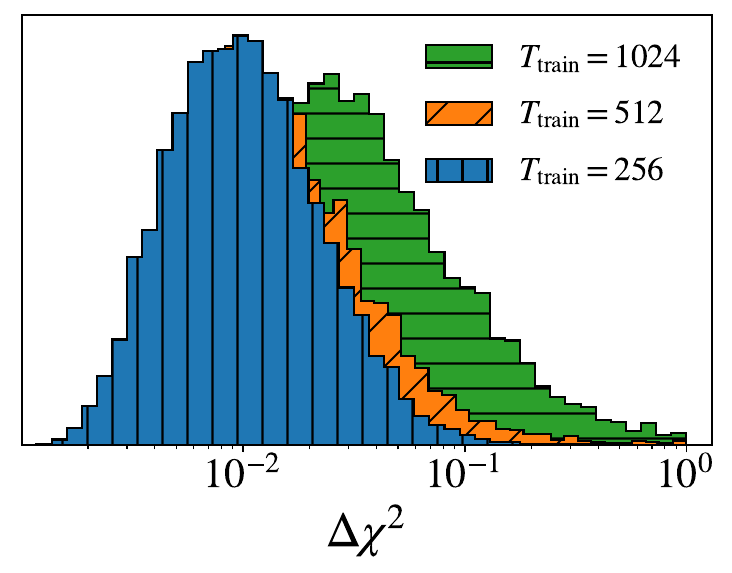}
    \caption{Histograms displaying the $\Delta\chi^2$ distribution resulting from optical weak lensing $3\times 2$pt forecasts for LSST Y1 for training temperatures of $T_{\rm train} = 1024$, $512$, and $256$. The values in the histogram represent evaluation of $10^4$ cosmologies on a testing set with $T_{\rm test}= T_{\rm train}/2$. The $T_{\rm train}= 512$ and $256$ distributions are both centered around $\Delta\chi^2 = 10^{-2}$ and have small tails that extend beyond $\Delta\chi^2=0.1$, reflecting the results in~\ref{tab:3x2pt_results}. The $T_{\rm train}=1024$ distribution is centered closer to $\Delta\chi^2=0.1$ with a tail that extends beyond $\Delta\chi^2=1$. Even with a loss function that does not penalize outliers, we see that the tails are not extended far into higher values of $\Delta\chi^2$. Furthermore, we see that lower temperatures do not necessarily improve in overall performance. This is likely caused by small gradients in the loss function near the minimum, meaning the model does not have the ability to fit the data more accurately. It is seen, however, that the tails of the distribution are under better control with lower training temperatures.}
    \label{fig:3x2pt_histograms}
\end{figure}

\begin{table}
    \centering
    \begin{tabular}{l|cccc}
        \hline
        $T_{\rm train}$                           & $128$   & $256$   & $512$   & $1024$ \\ \hline
        $\langle \Delta \chi^2 \rangle_{\rm med}$ & $0.006$ & $0.010$ & $0.012$ & $0.033$ \\
        \texthtbardotlessj$(\Delta \chi^2 > 0.2)$ & $<0.001$ & $0.001$ & $0.012$ & $0.074$ \\
        \hline
    \end{tabular}
    \caption{Summary of the results of the $3\times2$ point emulator, showing the median $\Delta\chi^2$ in the second row and the fraction of testing points with $\Delta\chi^2>0.2$ in the last row. All temperatures we consider are able to pass our threshold of having \texthtbardotlessj$(\Delta \chi^2 > 0.2) < 0.1$. Most prominently, even with a temperature as high as $128$, we are able to have zero testing points with $\Delta\chi^2>0.2$. In all cases, the median $\Delta\chi^2$ is well under control, being smaller than all values quoted in part II.}
    \label{tab:3x2pt_results}
\end{table}

\begin{figure*}[t]
    \centering
    \includegraphics[width=\textwidth]{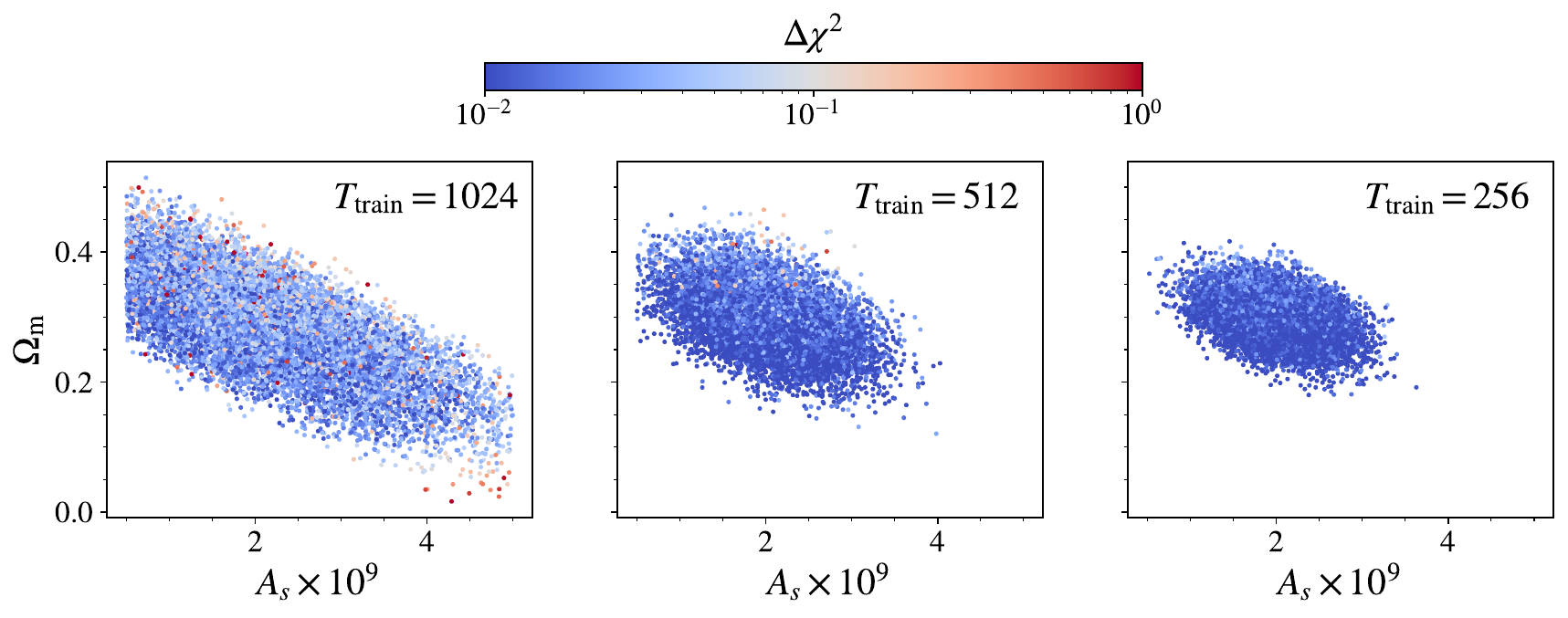}
    \caption{The projected spatial distribution of $\Delta\chi^2$ emulator errors into the $A_s\times 10^{9}$-$\Omega_{\rm m}$ plane. Each model was tested with $10^4$ testing points from a distribution with $T_{\rm test} = T_{\rm train}/2$. Interestingly, we find a slight bias, where larger values of $\Omega_{\rm m}$ perform worse than the smaller values, a feature not present when doing the same training procedure on a correlated posterior. This bias results from the additional parameter space introduced by the decorrelation procedure, and appears at simultaneously large values of $\Omega_{\rm m}$ and $H_{0}$. Thus, the bias is absent at the smaller temperatures. Aside from the bias, we can see that the training data is better centered on the chosen fiducial cosmology, meaning our training data is a more faithful representation of our input cosmology.}
    \label{fig:3x2pt_heatmap}
\end{figure*}

We find that all temperatures we can test are able to pass our imposed threshold, with the fraction of testing points with $\Delta\chi^2>0.2$ being less than $0.1$. With the adoption of the new loss function (Equation~\ref{eq:L3}) and activation function (Equation~\ref{eq:activation_fcn}), we are able to outperform our results from part II with far fewer training points, now only needing $10^4$ training points. Attempting to populate a larger region of parameter space by decorrelating the posterior introduced a mild bias in the emulator performance, despite the better average performance. This can possibly be resolved by selecting a more complicated parameter normalization during the emulator training, although we leave this for future work. The bias is not present when training on correlated posteriors. We also find that the transformer greatly benefits from having a large number of batches, and seems to only outperform the ResMLP models when the batch size is small.

\section{Supernova distance emulator}
\label{sec:SN}

In this Appendix, we describe the emulator used for the supernovae (SNe) luminosity distances. Parameterizing the dark energy equation of state using a basis of principal components (PCs), we aim to create a training set of cosmological parameters and distances. Following the analysis in \cite{Mortonson:2008qy}, we write
\begin{equation}
    w(z_j) = w_{\rm fid} + \sum_{i=1}^{N_{z,PC}} \alpha_i \mathbf{e}_i(z_j).
    \label{eq:w(z)}
\end{equation}
For numerical stability, we will normalize our basis vectors such that, 
\begin{equation}
    \sum_{i=1}^{N_{z,PC}}[\mathbf{e}_i(z_j)]^2 = \sum_{j=1}^{N_{z,PC}} [\mathbf{e}_i(z_j)]^2 = N_{z,PC}.
\end{equation}
We study smooth dark energy models that produce deviations from $\Lambda$CDM, which we assume is our fiducial model, and where $w_{fid}=-1$. We construct the Fisher matrix for the average magnitude of the SNe, and then we rotate to the basis where it is diagonal, 
\begin{equation}
    F_{ij}^{SN} = \sum_{\alpha} \sigma^{-2}_{\alpha} \frac{dm(z_{\alpha})}{d\theta_i} \frac{dm(z_{\alpha})}{d \theta_j},
    \label{eq:Fisher_mat}
\end{equation}
where $m(z_{\alpha}) = 5 \rm log[H_0d_L(z_{\alpha})] + \mathcal{M}$ is the average apparent magnitude of the SNe in the redshift bin denoted by $z_{\alpha}$, $\sigma_{\alpha}$ is the error in the average magnitude given in Appendix A of \cite{Mortonson:2008qy} and $\mathcal{M} = M - 5 \log(H_0/Mpc^{-1})+25$ is a constant related to the unknown absolute magnitude of the SNe. Within the parametrization that we have chosen, the parameters that are relevant are,
\begin{equation}
    \boldsymbol{\theta} = \{ \alpha_1, ... , \alpha_{N_{z,PC}}, \Omega_m, \Omega_m h^2\}
\end{equation}
where $ h = H_0 / (100 {\rm km \, s^{-1} Mpc^{-1}} )$. 

\begin{figure}
    \centering
    \includegraphics[width=0.9\columnwidth]{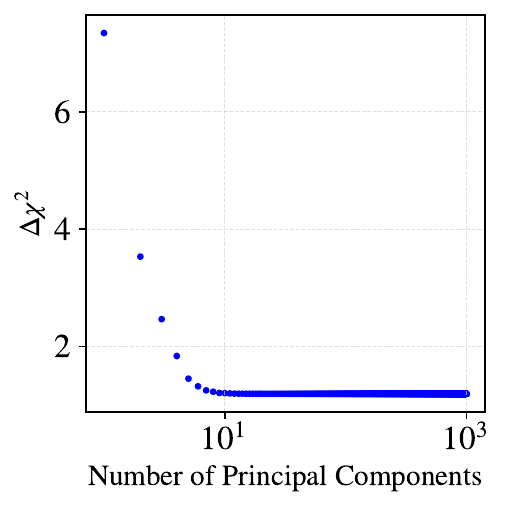}
    \caption{$\Delta \chi^2$ between the average magnitude $m(z)$ reconstructed using the Principal Component basis and the average magnitude $m(z)$ calculated analytically, as a function of the number of Principal Components used in the reconstruction for given cosmological parameters $\boldsymbol{\theta}$.}
    \label{fig:chiSq_vs_PCs}
\end{figure}

Our goal is to use this parametrization of $w(z)$ to generate a training set of cosmological inputs $\boldsymbol{\theta}$ with their corresponding luminosity distances $d_L(z)$, which will be used to train the emulator. Before proceeding, we first study the completeness of the PC basis that we use. Our goal is to determine the number of PCs we need to retain in order to construct $d_L(z)$ within the acceptable uncertainty. For the decomposition in \ref{eq:w(z)}, we use $N_{z,PC} = 1000$ PCs. In order to determine how many of these PCs we need to use to train the emulator, we compare the reconstructed $m(z)$ with the corresponding analytic expression. We find that retaining additional PCs beyond $N_{z,PC} \sim 20$ leads to minimal changes in the computed $\Delta \chi^2$. This result is illustrated in Figure~\ref{fig:chiSq_vs_PCs}, where we plot $\Delta \chi^2$ values as a function of the number of PCs used to reconstruct $m(z)$. Note that for this analysis, we have ranked the PCs in order of increasing uncertainty. To remain conservative, and since our approach does not involve computationally restrictive analytic calculations, we opt to retain 50 PCs. Using this set-up, we generate a training set of $\boldsymbol{\theta}$ and their corresponding $d_L(z)$. For the range of coefficients $\alpha_i$ sampled in the training set, we follow the constraints outlined in \cite{Mortonson:2008qy}.

\begin{figure}
    \centering
    \includegraphics[width=0.9\columnwidth]{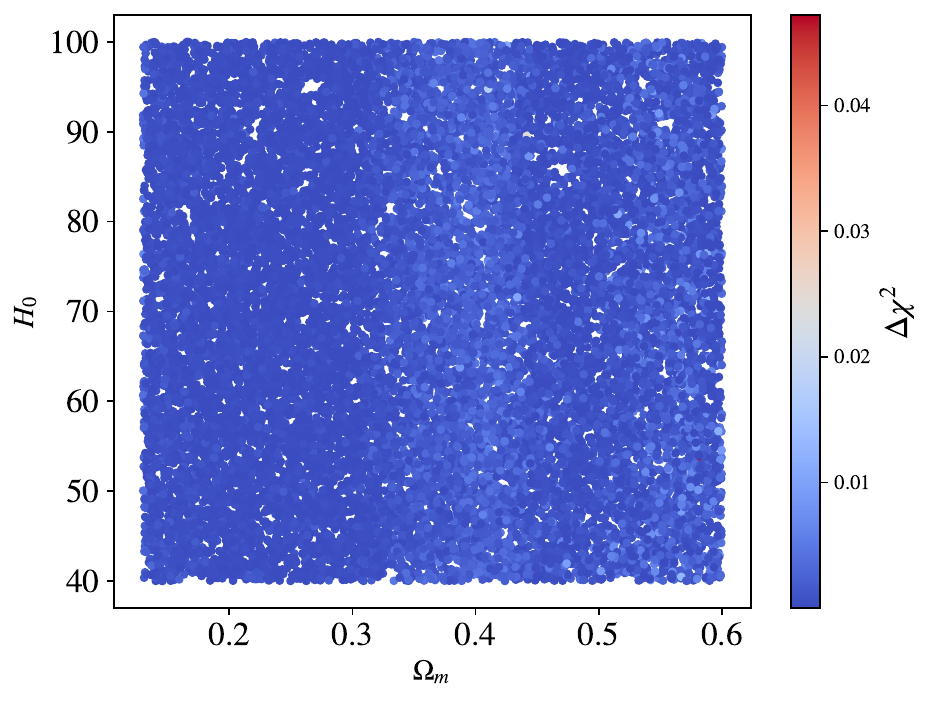}
    \caption{Heatmap of $\Delta \chi^2$ values comparing the emulator's $\log \textrm{H}_0 d_L(z)$ predictions to the true $\log \textrm{H}_0 d_L(z)$ from the test dataset. The plot shows $\Delta \chi^2$ as a function of $\Omega_m$ and $H_0$, with PC amplitudes implicit in order to maintain clarity.}
    \label{fig:chiSq_heatmap_50PCs}
\end{figure}

For the emulator, we find that the ResMLP architecture performs sufficiently well. The neural network is designed with an input layer consisting of 52 cosmological parameters (50 PC amplitudes, $\Omega_m$, $H_0$) and includes 8 ResBlocks. We use the Adam optimizer and the average $\Delta \chi^2$ over all z bins as the loss function during training, with the error definitions for each redshift bin provided in Appendix A of \cite{Mortonson:2008qy}. The training, validation, and testing datasets are generated by uniformly sampling the allowed ranges of the cosmological parameters $\Omega_m$, $H_0$, and PC amplitudes. Figure~\ref{fig:chiSq_heatmap_50PCs} shows the $\Delta \chi^2$ values comparing the emulator's predictions with the known luminosity distance $d_L(z)$ for the cosmologies in the test set. 

\section{1D Convolutional Neural Network}
\label{sec:cnn}

\begin{figure}[!ht]
    \centering
    \includegraphics[width=0.95\columnwidth]{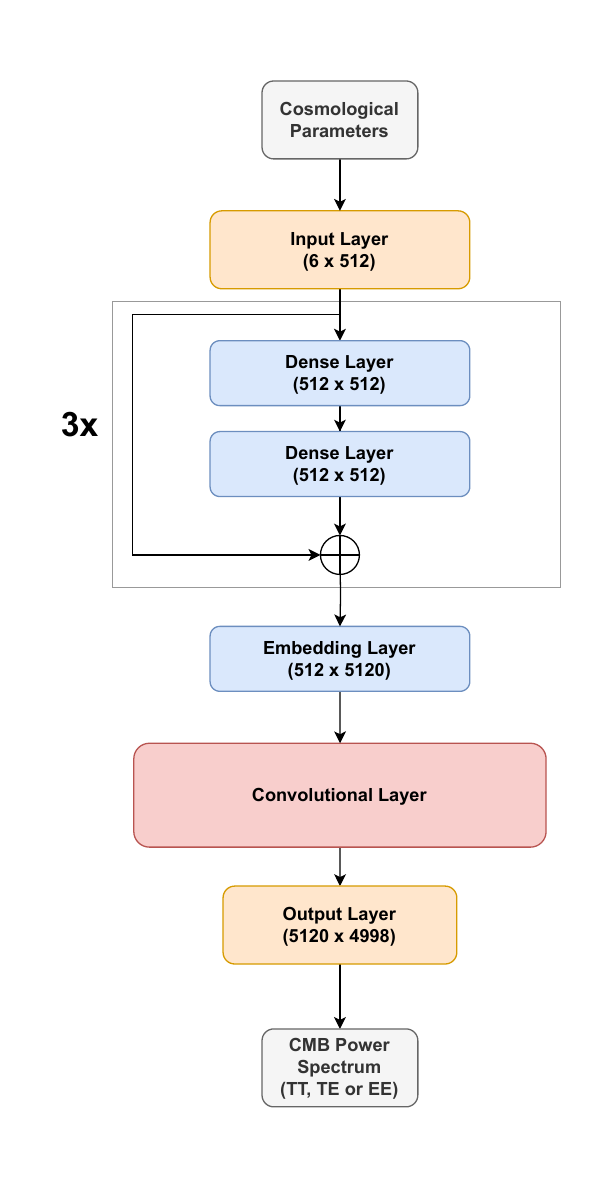}
    \caption{Architecture of the 1D convolutional neural network we use. Similar to the case with the transformer, we first pass the data through three ResBlocks. We then replace the attention and transformer with a convolutional layer. The convolutional layer contains \textsc{Kernel-Size}$=5$, \textsc{Stride}$=16$, \textsc{Padding}$=2$, \textsc{Out-Channel}$=16$.}
    \label{fig:CNN}
\end{figure}

There are additional, non-attention-based architectures that are worth exploring. A popular architecture used in image analysis is a convolutional neural network (CNN)~\cite{2019arXiv190503554K}. These models work by convolving an image with a kernel matrix. Since our data is vector-like, we instead use a convolution vector, creating a 1-dimensional CNN (1D-CNN)~\cite{Pandya:2025ueu}. We believe that this architecture can allow us to capture the smooth, short-range correlations in the CMB power spectra. We test one architecture in which we append a 1D-CNN layer without pooling after $3$ ResBlocks, as shown in Figure~\ref{fig:CNN}. The convolutional layer is set to have \textsc{Kernel-Size}$=5$, \textsc{Stride}$=16$, \textsc{Padding}$=2$, \textsc{Out-Channel}$=16$. The choice of \textsc{Out-Channel} is analogous to $N_{\rm channel}$ of transformers in the sense that the data vectors are rearranged into channels while the model processes the data vectors in pieces. Similarly, we want to move the kernel with the same amount of steps during convolution, so we have the same number setting for \textsc{Stride}. We choose one common \textsc{Kernel-Size} number for this primary test, and the setting for \textsc{Padding} is then chosen to give us the correct output dimension once \textsc{Kernel-Size} is fixed. 

For CMB power spectrum, we train it on $N_{\rm train}=530$ thousand data vectors with $T_{\rm train}=256$ training set, and test it on $T_{\rm test}=128$ testing set, which returns a \texthtbardotlessj$(\Delta \chi^2 > 0.2)=0.059$. This is roughly the same as the transformer architecture result, which has  \texthtbardotlessj$(\Delta \chi^2 > 0.2)=0.061$. CNN has the advantage of being faster to train since it does not have many trainable parameters.
\begin{figure}[t]
    \centering
    \includegraphics[width=\columnwidth]{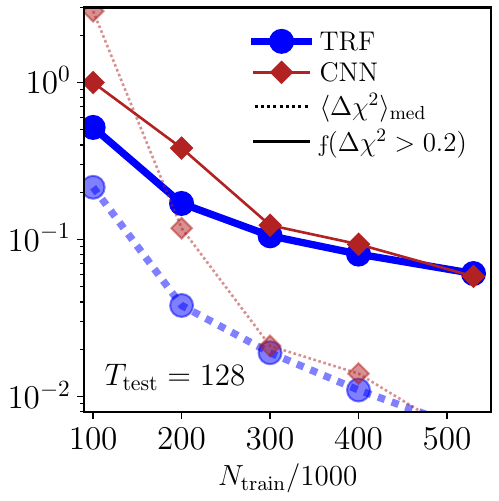}
    \caption{A comparison of the accuracy of a transformer and a 1D-CNN. The solid lines are for fraction of points with $\Delta\chi^2>0.2$, and the dashed lines are for median $\Delta\chi^2$. The thick blue lines are for transformers, and the thin red lines are for 1D-CNN. We did not optimize the hyperparameters for the 1D-CNN. With about $1\times10^5$ training points, the transformer outperforms the 1D-CNN by about a factor of two in \texthtbardotlessj$(\Delta\chi^2>0.2)$, and by about an order of magnitude in $\langle \Delta\chi^2\rangle_{\rm med}$. However, as the number of training points increases to about $3\times10^{5}$ and beyond, the performance of the transformer and 1D-CNN are approximately equal.}
    \label{fig:cnnvstrf}
\end{figure}

\section{Improved Modeling of the Damping Tail and Lensing}\label{app:rescale}

The presence of a damping tail in the CMB power spectra leads to additional difficulties in training machine learning models. By pre-processing the TT power spectrum using the techniques in section~\ref{sec:train_strat_cmb}, we are able to reduce the span of the data vectors by orders of magnitude, which assists in training the emulator. To improve this, we deploy Symbolic Regression to develop a fit of the TT power spectrum damping/lensing tail.

Genetic Algorithms (GAs) are a class of machine learning techniques that learn to fit data by mimicking evolutionary biology, such as random gene mutation and crossbreeding~\cite{Koza_1992}. GAs have been widely used in the field of cosmology, for example, to model the linear matter power spectrum~\cite{orjuelaquintana2024analyticalemulatorlinearmatter}, nonlinear matter power spectrum\cite{Bartlett:2024jes}, and matter transfer function~\cite{Orjuela_Quintana_2023}. Symbolic Regression, in particular, is a kind of GA that attempts to learn a symbolic expression to fit a given dataset.

We design a GA model that takes a fixed functional form for the lensed damping tail. The functional form for our fit has two components, the damping tail and the lensing effect, and a total of $15$ free parameters, $\alpha_i$. The lensing contribution is described in Equation.~\ref{eq:SR_template}. For the damping tail, we follow the work done in \cite{Hu:1996mn}. There, the rescaling factor, to go from the undamped power spectrum without lensing tail to the full damped unlensed power spectrum, is given by
\begin{equation}\label{eq:damping_tail}
T_D(\ell)=\frac{A_s}{e^{2\tau}}\mathscr{D}_\ell^2 \mathscr{R}_\ell^2 \mathscr{P}_\ell,
\end{equation}
where $\mathscr{D}_\ell$ is the diffusion damping contribution, $\mathscr{R}_\ell$ is the reionization damping contribution, and $\mathscr{P}_\ell$ is the potential envelope damping contribution. The diffusion-damping contribution is modeled as
\begin{equation}
    \mathscr{D}_\ell = e^{-(\ell/\ell_D)^m}
\end{equation}
where $\ell_D$ and $m$ are given by
\begin{align} \label{eqn:damping_diff_scale}
\ell_D/r_\theta^\star &= a_1\left(\Omega_b h^2\right)^{0.291} \left[1+a_2\left(\Omega_b h^2\right)^{1.80}\right]^{-1/5}\\
m&=a_3\left(\Omega_b h^2\right)^{a_4} \left[1+\left(\Omega_b h^2\right)^{1.80}\right]^{1/5}
\end{align}
where $r_\theta^\star$ is the comoving angular-size distance to last scattering. The $a_i$ are power law functions of $\Omega_mh^2$, given by
\begin{align}
    a_1 &= \alpha_1\left(\Omega_mh^2\right)^{\alpha_2}\left[1+\alpha_3\left(\Omega_mh^2\right)^{\alpha_4}\right],\\
    a_2 &= \alpha_5\left(\Omega_mh^2\right)^{\alpha_6}\left[1+\alpha_7\left(\Omega_mh^2\right)^{\alpha_8}\right]^{-1},\\
    a_3 &= \alpha_9\left(\Omega_mh^2\right)^{\alpha_{10}},\\
    a_4 &= \alpha_{11}\left(\Omega_mh^2\right)^{\alpha_{12}}.
\end{align}

The reionization damping term is given by
\begin{equation}
\mathscr{R}_{\ell}^2 = \frac{1-e^{-2\tau}}{1 + c_1x + c_2x^2 + c_3x^3 + c_4 x^4} + e^{-2\tau},
\end{equation}
with $x=\ell/\left(\ell_r + 1\right)$ and constants $c_1 = -0.276$, $c_2 = 0.581$, $c_3 = -0.172$, and $c_4 = 0.0312$. The parameter $\ell_r=\left(\eta_0 - \eta_r\right)/\eta_r$, where $\eta_r$ is the comoving distance to reionization. In this study, we set the redshift when reionization occurs at $z=10$.

The potential envelope damping contribution is approximately given by
\begin{equation}
    \mathscr{P}_\ell = 1 + A\exp{(-1.4\ell_{eq}/\ell)},
\end{equation}
where $\ell_{eq}=k_{eq}r_\theta^\star$, and $k_{eq}$ is the scale that crosses the horizon at matter-radiation equality. The amplitude $A$ is fixed by the expression
\begin{equation}\label{eq: A}
A=25\left(1+\frac{4}{15}f_\nu\right)^{-2}\frac{(1+R_\star)^{-1/2}+(1+R_\star)^{-3/2}}{2}-1,
\end{equation}
where $f_\nu \ = \rho_{\nu}/(\rho_{\nu} + \rho_{\gamma})$ is the fraction of the radiation energy density comprised of neutrinos, and $R_* = \frac{3}{4}\rho_{b}(\eta_*)/ \rho_{\gamma}(\eta_*)$ is the ratio of baryon and photon densities at matter-radiation equality.

\begin{figure}[t]
    \centering
    \includegraphics[width=\columnwidth]{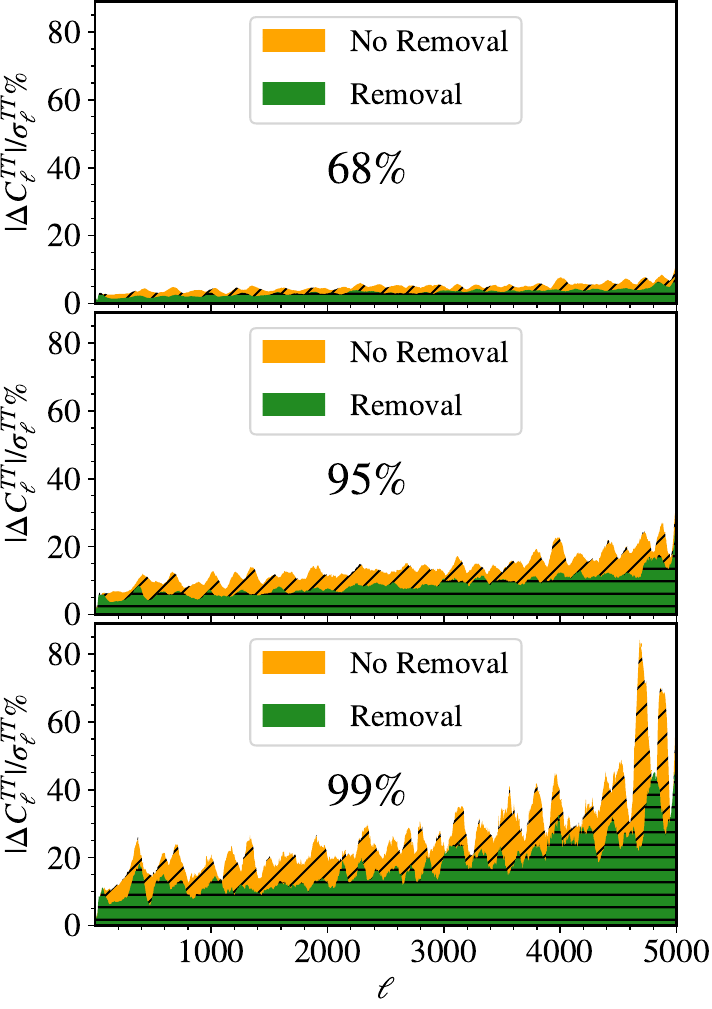}
    \caption{This plot shows the distributions of the ratio between the absolute error of the emulator and the cosmic variance standard deviation in percentage, with and without removal of the damping tail, both trained on the same architecture and training settings. We can see that we get better emulation on high $\ell$ region with the damping tail removal, which helps us to get a factor of $2$ improvement in $\langle\Delta\chi^2\rangle_{\rm med}$. }
    \label{fig:GAcomp}
\end{figure}

\begin{table}[t]
    \centering
    \begin{tabular}{c c | c c } 
     \hline\hline
     Parameter & Value & Parameter & Value \\ 
     \hline\hline
     $\alpha_1$ & 0.082 &$\alpha_7$ & 1.3 \\ 
     \hline
     $\alpha_2$ & 0.035 & $\alpha_8$ & -0.11 \\
     \hline
     $\alpha_3$ & 5.0 & $\alpha_9$ & 1.3 \\
     \hline
     $\alpha_4$ & 0.47 & $\alpha_{10}$ & 0.022  \\
     \hline
     $\alpha_5$ & 1300 &$\alpha_{11}$ & 0.059  \\ 
     \hline
     $\alpha_6$ & 0.46  & $\alpha_{12}$ & 0.17\\ 
    \end{tabular}
    \caption{The GA learned parameters of the damping tail fit as described by equations \ref{eq:damping_tail} - \ref{eq: A}.}
    \label{GAlearned}
\end{table}

Thus constructed, Equations~\ref{eq:damping_tail} through \ref{eq: A} comprise the functional form for the damping tail fit of the TT power spectrum we consider. The learned parameters are listed in Table~\ref{GAlearned}.

For the lensing effect, we employ Symbolic Regression (SR). This technique works by randomly generating a population of expression trees, known as candidate models, and evolving them over a number of evolutionary steps. At each step, either one fit expression is chosen from the population and randomly mutated to produce a child expression, or two fit expressions are selected as parents and mixed, or crossed-over, to produce two child expressions. Whichever the selected operation, the resulting expression(s) replace the oldest expression(s) in the population. As this process repeats, the fittest expressions tend to survive as the overall population of candidate expressions evolves to fit the data. By the end of learning, the fittest expression in the final population is returned.

To perform SR, we use the Python package PySR, an open-source Python library for SR built on a highly optimized Julia backend. The core PySR algorithms are described in \cite{cranmer2023interpretablemachinelearningscience}. PySR performs SR on independent populations of expressions asynchronously, and allows for expressions to migrate across populations after a set number of evolutionary steps, leading to efficient and parallelizable learning.

PySR's inner loop is a simple evolutionary algorithm performed on each population and constitutes one evolutionary step. Firstly, an evolution is chosen from the following: a selected expression is randomly mutated, two fit expressions are selected to be crossed-over, a selected expression is algebraically simplified, or a selected expression's constants are optimized. To select an expression for evolution, or two if crossing-over, a random subset of the population is chosen, and the best fit expression(s) within, according to a specified fitness function, is selected. If child expressions are produced, via random mutation or crossing-over, these then replace the oldest expressions in the population. If simplification or optimization of the selected expression is chosen, the algorithm directly modifies the original expression in memory.

The outer loop in PySR allows the most fit expressions in each population to migrate across populations in an effort to mitigate the risk that the isolated populations get stuck in a local minimum of the search space. As populations are evolving, PySR stores the best expressions it has seen by population. After all populations have gone through a set number of evolutions, the best seen expressions have a chance at being randomly inserted back into other populations. Following migration, the populations are evolved independently again. The outer loop constitutes one iteration of learning and runs for however many iterations the user specifies.

Besides featuring a robust and optimized implementation of SR, PySR is also highly configurable. PySR allows the user to define a template expression, within which SR learning takes place in a structured way. The user fixes the outer form of the expression and specifies which inner expression(s) and parameters(s) PySR may learn. We take a power-law function as the template expression for our fitting formula of the lensing tail:

\begin{equation}
    L\left(\ell\right) = 1 + w\left(\ell \right)\left[\alpha_{13}\left(\frac{\ell}{\alpha_{14}}\right)^{\alpha_{\rm{SR}}\left(\Omega_bh^2,\  \Omega_mh^2\right)} - 1\right],\label{eq:SR_template}
\end{equation}
where
\begin{equation}
w\left(\ell\right) = \frac{1}{1+e^{-(\ell-\alpha_{15})/100}}
\end{equation}
is a sigmoid function, included to smoothly transition into the lensing tail. The width of the sigmoid is fixed to 100. The parameters $\alpha_i$ are to be tuned by PySR in the process of learning. $\alpha_1$ and $\alpha_2$ are added to allow for some flexibility in the scaling of the power-law, and $\alpha_3$ allows the midpoint of the sigmoid to be learned. $\alpha_{SR}$ is the power-law exponent which we leave for PySR to learn an analytic expression for in terms of $\Omega_bh^2$ and $\Omega_mh^2$.

We set PySR to learn for $200$ iterations with $20$ populations of 1000 candidate expressions. We leave the fitness function as the PySR default, which is the mean-square error. We restrict PySR's set of possible operators in constructing expressions to be addition, subtraction, multiplication, and exponentiation. The maximum complexity of the expressions, which limits the number of nodes possible in an expression tree, is set to $30$. All other settings and hyperparameters are left at their PySR default values.

After learning, PySR converged to the following values for $\alpha_i$:
\begin{align}
    \alpha_{13} &= 0.83 \notag\\
    \alpha_{14} &= 3218 \notag\\
    \alpha_{15} &= 3240 \notag.
\end{align}
The learned exponent is
\begin{equation}
    \alpha_{\rm{SR}} = \Omega_mh^2\left(\Omega_bh^2\right)^{-0.877}-3.342\Omega_mh^2-1.118.\label{eq:SR_exponent}
\end{equation}

\begin{figure}[t]
    \centering
    \includegraphics[width=\columnwidth]{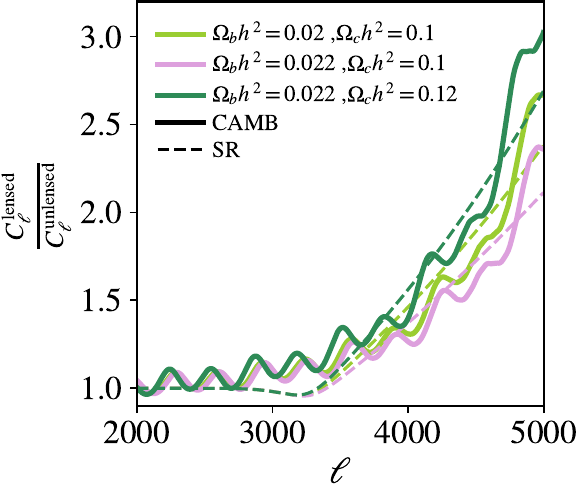}
    \caption{This plot shows the comparison of the lensing tail ratio between lensed and unlensed CMB TT power spectrum computed from \textsc{CAMB} and from the analytical approximation trained with PySR, on $3$ cosmologies. We can see the SR recovers the lensing tail up to the same order of magnitude.}
    \label{fig:SR}
\end{figure}

Thus constructed, Equations~\ref{eq:SR_template} through \ref{eq:SR_exponent} comprise the functional form for the lensing tail fit of the TT power spectrum we consider. We show how well we recover the rough shape and order of magnitude of the lensing tail in Figure~\ref{fig:SR}.

As a preliminary test, we emulate the CMB TT power spectrum assuming $\Lambda$CDM  with three free parameters: $\Omega_bh^2, \Omega_ch^2$ and $H_0$. We employ a ResMLP architecture with PCA decomposition of the data vectors. We use $1.8 \times 10^4$ data vectors for training. The parameter value ranges in which we train and test the models are listed in Table~\ref{GArange}. The results of removing and not removing the damping tail are in Table~\ref{tab:GAresult}, and a comparison between the distribution of ratios between the absolute error of the emulator and the cosmic variance standard deviation in percentage is shown in Figure~\ref{fig:GAcomp}. 

The removal of lensed damping tail provides an improvement of about a factor of $2$ in both $\langle\Delta\chi^2\rangle_{\rm med}$ and \texthtbardotlessj($\Delta\chi^2>10$) with our fitting results. We leave a more comprehensive analysis, including the extension to a larger parameter volume box, the inclusion of additional parameters, and the adoption of alternative analytical expressions, to future work. Here, we highlight that incorporating a more accurate pre-processing step to account for the effects of damping could potentially significantly improve emulator performance. 

\begin{table}[H]
    \centering
    \renewcommand{\arraystretch}{1.3}
    \begin{tabular}{lcc}\hline
        Parameter & Remove-Damp Train & Remove-Damp Test \\\hline
        $100\Omega_b h^2$     & $[1.5,3.3]$ & $[1.6,3.2]$  \\
        $10\Omega_c h^2$     & $[0.3,1.8]$  & $[0.4,1.7]$ \\
        $H_0$              & $[45,95]$      & $[50,90]$       \\\hline
    \end{tabular}%
    \caption{The parameter limits we use in testing the removal of lensed damping tail. We only vary three parameters within a relatively smaller range than in Table~\ref{tab:parameter_ranges}.}
    \label{GArange}
\end{table}

\begin{table}[H]
     \centering
     \renewcommand{\arraystretch}{1.3}
     \begin{tabular}{l|c}\hline
          ResMLP          & $N_{\rm train}=18$k Train \\\hline
          No Removal & $10.1$ $(0.50)$ \\
          Removal     & $4.6$ $(0.23)$\\\hline
     \end{tabular}
     \caption{Comparison between ResMLP models trained with and without removal of lensed damping tail. The number outside of the parenthesis is the $\langle\Delta\chi^2\rangle_{\rm med}$ and the number inside the parenthesis is the fraction of outlier \texthtbardotlessj($\Delta\chi^2>10$). We compare \texthtbardotlessj($\Delta\chi^2>10$) because those models haven't reached the level of precision as high as before.}\label{tab:GAresult}
\end{table}

\bibliographystyle{apsrev4-1}
\bibliography{ml}
\end{document}